\documentclass[structabstract]{aa}  
\usepackage{amssymb}
\usepackage{amsmath}
\usepackage{amsfonts}
\usepackage[latin1]{inputenc}
\usepackage{graphicx}
\usepackage{epsfig}
\usepackage{subfigure}
\usepackage[authoryear]{natbib}
\usepackage[figuresright]{rotating}
\usepackage{supertabular}

\usepackage{txfonts}
\newcommand{\teff}{\ifmmode T_{\rm eff} \else $T_{\mathrm{eff}}$\fi}
\newcommand{\logg}{\ifmmode \log g \else $\log g$\fi}
\newcommand{\lL}{\ifmmode \log \frac{L}{L_{\odot}} \else $\log \frac{L}{L_{\odot}}$\fi}
\newcommand{\mdot}{$\dot{M}$}

\newcommand{\vsini}{$V$ sin$i$}
\newcommand{\vinf}{$v_{\infty}$}

\newcommand{\kms}{km~s$^{-1}$}
\newcommand{\msun}{\ifmmode M_{\odot} \else $M_{\odot}$\fi}
\newcommand{\zsun}{\ifmmode Z_{\odot} \else $Z_{\odot}$\fi}
\newcommand{\lsun}{\ifmmode L_{\odot} \else $L_{\odot}$\fi}
\newcommand{\rsun}{\ifmmode R_{\odot} \else $R_{\odot}$\fi}
\newcommand{\qh}{\ifmmode Q_{\rm H} \else $Q_{\rm H}$\fi}
\newcommand{\qhei}{\ifmmode Q_{\ion{He}{i}} \else $Q_{\ion{He}{i}}$\fi}

\begin{document}
   \title{Evidence for a physically bound third component in HD\,150136}


   \author{L. Mahy\inst{1}
          \and
          E. Gosset\inst{1}
          \and
          H. Sana\inst{2}
          \and
          Y. Damerdji\inst{1}
          \and
          M. De Becker\inst{1}
          \and
          G. Rauw\inst{1}
          \and
          C. Nitschelm\inst{3}
          }

   \offprints{L. Mahy}

   \institute{Institut d'Astrophysique et de G\'eophysique, Universit\'e de Li\`ege, B\^at. B5C, All\'ee du 6 Ao\^ut 17, B-4000, Li\`ege, Belgium\\
              \email{mahy@astro.ulg.ac.be}
         \and 
             Sterrenkundig Instituut "Anton Pannekoek", University of Amsterdam,
Postbus 94249, NL-1090 GE Amsterdam, The Netherlands
         \and
             Instituto de Astronom{\'i}a, Universidad Cat{\'o}lica del Norte, Avenida Angamos 0610, Antofagasta, Chile
         }

   \date{Received ...; accepted ...}

 
  \abstract
   {HD\,150136 is one of the nearest systems harbouring an O3 star. Although this system was for a long time considered as binary, more recent investigations have suggested the possible existence of a third component.}
   {We present a detailed analysis of HD\,150136 to confirm the triple nature of this system. In addition, we investigate the physical properties of the individual components of this system.}
   {We analysed high-resolution, high signal-to-noise data collected through multi-epoch runs spread over ten years. We applied a disentangling program to refine the radial velocities and to obtain the individual spectra of each star. With the radial velocities, we computed the orbital solution of the inner system, and we describe the main properties of the orbit of the outer star such as the preliminary mass ratio, the eccentricity, and the orbital-period range. With the individual spectra, we determined the stellar parameters of each star by means of the CMFGEN atmosphere code.}
   {We offer clear evidence that HD\,150136 is a triple system composed of an O3V((f$^{*}$))--3.5V((f$^{+}$)), an O5.5--6V((f)), and an O6.5--7V((f)) star. The three stars are between 0--3 Myr old. We derive dynamical masses of about  64, 40, and 35~\msun\ for the primary, the secondary and the third components by assuming an inclination of $49\degr$ ($\sin^3{i} = 0.43$). It currently corresponds to one of the most massive systems in our galaxy. The third star moves with a period in the range of 2950 to 5500~d on an outer orbit with an eccentricity of at least 0.3. However, because of the long orbital period, our dataset is not sufficient to constrain the orbital solution of the tertiary component with high accuracy.}
   {We confirm the presence of a tertiary star in the spectrum of HD\,150136 and show that it is physically bound to the inner binary system. This discovery makes HD\,150136 the first confirmed triple system with an O3 primary star.}


   \maketitle


\section{Introduction}
\label{s_intro}
NGC\,6193 is a young open cluster surrounded by an \ion{H}{ii} region in the Ara\,OB1 association. The core of the cluster hosts two bright and massive stars separated by about $10 \arcsec$: HD\,150135 and HD\,150136. 

Several investigations in different wavelength domains (optical, radio, and X-rays) allowed to improve the knowledge of HD\,150136. In the optical domain, \citet{ng05} reported this star as a binary system composed of an O3 primary and an O6 secondary. Indeed, spectral absorption lines of \ion{N}{iv}~$\lambda{5203}$ and \ion{N}{v}~$\lambda \lambda{4603-19}$ (wavelengths are hereafter expressed in \AA), as well as the emission line \ion{N}{iv}~$\lambda{4058}$, were observed in the composite spectra. These lines are considered to be typical features for O2--O3 stars \citep{wal02}. This is a rather rare category in which only a few of these stars are reported as binary systems. Therefore, even though these objects are considered, with the WNLh stars, as possibly belonging to the most massive objects, their actual masses are scarcely known. 

Located at a distance of $1.32\pm0.12$~kpc \citep{hh77}, HD\,150136 would thus harbour the nearest O3 star and would constitute a target of choice for investigating the fundamental parameters of such a star. Initially, the orbital period of the system was estimated to about 2.7~d whilst the mass ratio between the primary and the secondary stars was measured to be equal to 1.8 (unpublished, see \citealt{gar80}). The analysis of \citet{ng05} allowed the authors to refine these values to 2.662~d for the orbital period and 1.48 for the mass ratio. Moreover, the possibility of a third component bound to the system was envisaged (though no clear feature was detected) to explain that the \ion{He}{i} lines in the spectra did not follow the orbital motion of any of the binary components.

In the radio domain, HD\,150136 was reported as a non-thermal radio emitter \citep{ben06,deb07}, which suggests the presence of relativistic electrons accelerated in shocks issued from a wind-collision region. This non-thermal source could be of about $2\arcsec$ \citep[i.e., the resolution,][]{ben06} and could be associated with a close visual component separated from HD\,150136 by about $1.6\arcsec$ \citep{mas98}, i.e., about 0.01~pc if we assume a distance of about 1.3~kpc. So far, there has however been no indication that this star is physically bound to the binary system.

The {\it Chandra} X-ray observations of the young open cluster NGC\,6193 revealed that HD\,150136 is one of the X-ray brightest O-type stars known \citep[$\log L_{\mathrm{X}}$ (erg~s$^{-1}$) = 33.39 and $L_{\mathrm{X}}/L_{\mathrm{bol}}=10^{-6.4}$, ][]{ski05}. Its emission probably results from a radiative colliding wind interaction and appears to be slightly variable on a timescale smaller than one day. According to \citet{ski05}, the discrepancy between the orbital period and the timescale of the variation in this X-ray emission is probably an occultation effect either linked to the stellar rotation or to a putative third star with a rather short period.

All these investigations thus suggest that HD\,150136 is a multiple system rather than a binary one, even though evidence is still lacking. In this context, an intense spectroscopic monitoring at high resolution in the optical domain, first initiated to study the binary itself, was performed to bring new light on this system. We organize the present paper as follows. Section~\ref{s_obs} describes the observations and the data reduction technique. Section~\ref{s_dis} reveals the existence of a third star in the data of HD\,150136 and focuses on the disentangling process in order to compute the individual spectra of the components. A first determination of the orbital solution of the known short-period system based on an SB2 configuration is presented in Sect.~\ref{s_inner}. We also demonstrate that these parameters can be optimized by assuming the external influence of a third component. The possible orbit of this third star is addressed in Sect.~\ref{s_outer} and an SB3 partial solution is derived. This discovery makes HD\,150136 one of the first systems where it is demonstrated that the third component is physically bound to the inner system (see also Tr16-104, \citealt{rau01}, and HD\,167971, \citealt{blo07}). However, it is the first time that the primary is an O\,3 star. We thus in Sect.~\ref{s_spec} use the CMFGEN atmosphere code \citep{hm98} to derive the fundamental properties of the stars to improve our knowledge on the O\,3 star parameters. Moreover, the configuration of the short-period binary seems to suggest a wind interaction zone possibly detectable through the \ion{He}{ii}~$\lambda{4686}$ and H$\alpha$ lines. We devote Sect.~\ref{s_tomo} to the analysis of the variability observed in these lines by applying a Doppler tomography technique. Finally, we discuss the implications of our results in Sect.~\ref{s_disc} and we provide the conclusions in Sect.~\ref{s_conc}.


\section{Observations and data reduction}
\label{s_obs}
A multi-epoch campaign was devoted to observing HD\,150136. Between 1999 and 2009, we collected 79 high-resolution, high signal-to-noise spectra of this star with the \'echelle Fiber-fed Extended Range Optical Spectrograph (FEROS) successively mounted at the ESO-1.5m (for spectra taken in 1999 and 2001) and the ESO/MPG-2.2m (for observations between 2002 and 2009) telescopes at La Silla (Chile). This instrument has a resolving power of about 48000 and the detector was a 2k $\times$ 4k EEV CCD with a pixel size of 15$\mu$m $\times$ 15$\mu$m. Exposure times were between 2 and 10~min to yield a signal-to-noise ratio over 200, measured in the continuum on the region [$4800-4825$]~\AA. FEROS is providing 39 orders allowing the entire optical wavelength domain, going from 3800 to 9200~\AA, to be covered. The journal of the observations is listed in Table~\ref{tab1} (including the individual signal-to-noise ratios). For data reduction process, we used an improved version of the FEROS pipeline working under the MIDAS environment \citep{san06a}. The data normalization was performed by fitting polynomials of degree $4-5$ to carefully chosen continuum windows. We mainly worked on the individual orders, but the regions around the \ion{Si}{iv}~$\lambda \lambda{4089}$,~$4116$, \ion{He}{ii}~$\lambda{4686}$, and H$\alpha$ emission lines were normalized on the merged spectrum. We also note that the flatfields of the 2009 campaign presented some oscillations of unknown origin. These oscillations represent about $3$\% to $5$\% of the continuum flux. We applied a filter to smooth these flatfields and thus to limit the effects of the undesirable oscillations on the observed spectra.  

\begin{table*}
\caption{Journal of observations.} \label{tab1}      
\centering            
\begin{tabular}{c c r r r | c c r r r}
\hline\hline 
HJD & $S/N$ & $RV_{\mathrm{P}}$ & $RV_{\mathrm{S}}$ & $RV_{\mathrm{T}}$ & HJD & $S/N$ & $RV_{\mathrm{P}}$ & $RV_{\mathrm{S}}$ & $RV_{\mathrm{T}}$\\
$- 2\,450\,000$&  &  [\kms]   &  [\kms]  & [\kms] & $-2\,450\,000$&  &  [\kms]   &  [\kms]  & [\kms]\\
\hline                                 
1327.9031 &311  &  173.5 & $-$368.8 &  25.4 & 3511.8816 &308  & $-$182.1 &  247.7 & $-$54.6\\ 
2037.7746 &214  & $-$234.8 &  263.1 &  14.3 & 3512.5377 &243  &  114.6 & $-$217.9 & $-$48.5\\
2037.9247 &231  & $-$246.8 &  292.5 &  13.2 & 3512.8633 &344  &  193.7 & $-$344.7 & $-$53.0\\
2381.6519 &384  &  167.6 & $-$355.2 &  $-$6.4 & 3512.8663 &281  &  193.7 & $-$347.6 & $-$50.6\\ 
2381.8167 &404  &  156.4 & $-$314.5 &   0.2 & 3796.7553 &210  &  152.1 & $-$255.5 & $-$70.3\\ 
2382.6532 &346  & $-$184.6 &  232.9 &  $-$2.3 & 3796.7576 &204  &  154.0 & $-$251.6 & $-$70.4\\ 
2382.7995 &380  & $-$222.9 &  280.8 &  $-$0.7 & 3797.7890 &219  & $-$211.3 &  334.1 & $-$72.4\\ 
2383.6582 &339  &   16.4 &  $-$71.8 &  $-$3.0 & 3797.7931 &207  & $-$212.5 &  331.3 & $-$72.9\\ 
2383.8118 &434  &   84.2 & $-$158.7 &   5.8 & 3797.9131 &271  & $-$206.0 &  318.1 & $-$70.1\\ 
2782.6246 &538  &  172.7 & $-$332.9 & $-$19.3 & 3797.9161 &307  & $-$203.6 &  317.8 & $-$71.7\\ 
2783.6661 &461  & $-$139.0 &  152.8 & $-$18.4 & 3798.7598 &294  &  131.4 & $-$229.7 & $-$66.2\\ 
2784.6897 &389  &  $-$70.9 &   35.4 & $-$18.3 & 3798.7628 &243  &  130.5 & $-$232.6 & $-$68.3\\ 
3130.6387 &399  &  167.7 & $-$329.1 & $-$33.9 & 3798.9088 &376  &  173.4 & $-$302.5 & $-$69.3\\ 
3130.6407 &355  &  169.1 & $-$325.8 & $-$32.5 & 3799.9158 &241  &  $-$75.4 &   86.0 & $-$63.2\\
3131.6332 &374  & $-$216.6 &  292.9 & $-$28.1 & 3800.7716 &262  & $-$163.5 &  254.6 & $-$71.8\\
3131.6351 &443  & $-$211.1 &  296.6 & $-$27.6 & 3800.7746 &308  & $-$163.4 &  249.1 & $-$70.9\\ 
3131.9266 &391  & $-$215.6 &  306.8 & $-$26.3 & 3800.9192 &374  & $-$113.9 &  156.9 & $-$74.1\\
3131.9286 &396  & $-$223.7 &  303.3 & $-$28.7 & 3861.6554 &278  & $-$158.9 &  239.1 & $-$72.8\\
3132.6081 &290  &   52.3 & $-$110.1 & $-$30.4 & 3861.9066 &339  & $-$210.5 &  324.1 & $-$68.1\\
3132.6101 &380  &   53.1 & $-$112.1 & $-$31.2 & 3862.8769 &322  &  113.4 & $-$171.7 & $-$69.6\\
3132.8874 &460  &  144.6 & $-$294.1 & $-$32.9 & 3863.6503 &357  &  147.8 & $-$243.0 & $-$69.2\\
3132.8893 &479  &  149.8 & $-$287.3 & $-$29.2 & 3863.8683 &336  &   51.0 &  $-$85.6 & $-$73.6\\
3133.6779 &418  &   48.9 & $-$109.2 & $-$38.2 & 3864.6371 &353  & $-$218.1 &  329.3 & $-$70.8\\
3133.6798 &451  &   52.5 & $-$107.3 & $-$36.1 & 3864.8531 &341  & $-$193.5 &  298.4 & $-$72.6\\
3133.9098 &376  &  $-$73.6 &   59.9 & $-$30.0 & 5021.4921 &578  &  172.2 & $-$356.7 &  10.9\\
3134.6256 &379  & $-$222.1 &  304.4 & $-$29.4 & 5021.5611 &527  &  157.0 & $-$342.4 &   9.2\\
3134.6275 &474  & $-$213.7 &  302.9 & $-$29.4 & 5021.6147 &519  &  145.3 & $-$315.3 &   7.9\\
3134.9164 &295  & $-$138.4 &  155.4 & $-$34.4 & 5022.4910 &461  & $-$211.5 &  257.4 &   9.9\\
3134.9183 &318  & $-$135.0 &  146.7 & $-$34.0 & 5022.7219 &548  & $-$252.6 &  306.6 &   8.9\\
3135.6251 &317  &  164.2 & $-$329.6 & $-$33.8 & 5023.4973 &378  &   29.2 & $-$117.8 &   9.1\\
3135.6271 &346  &  164.5 & $-$323.3 & $-$36.7 & 5023.7349 &419  &  118.2 & $-$278.0 &   9.4\\
3135.8725 &356  &  190.3 & $-$355.7 & $-$31.9 & 5025.5220 &386  & $-$236.7 &  283.7 &   7.9\\
3135.8744 &376  &  189.5 & $-$352.8 & $-$33.3 & 5025.7098 &474  & $-$188.9 &  211.0 &  10.4\\
3509.5753 &237  &  $-$28.0 &    1.3 & $-$49.2 & 5028.5145 &472  & $-$142.5 &  133.9  &  3.6\\
3509.5784 &220  &  $-$31.6 &   11.2 & $-$48.9 & 5028.7328 &342  &  $-$57.7 &  $-$15.7  &  8.2\\
3510.7732 &273  &   62.8 & $-$135.0 & $-$50.3 & 5029.4986 &455  &  167.5 & $-$354.3 &   7.9\\
3510.7762 &263  &   61.3 & $-$131.0 & $-$51.6 & 5029.7368 &330  &  120.2 & $-$269.5 &   3.5\\
3511.5647 &306  & $-$229.1 &  316.3 & $-$45.8 & 5031.4953 &487  &   10.9 &  $-$99.6 &  12.5\\
3511.5677 &316  & $-$229.3 &  319.7 & $-$45.9 & 5031.6685 &391  &   92.6 & $-$228.8 &   6.0\\
3511.8786 &300  & $-$186.6 &  249.9 & $-$53.8 & 	      &	    &	     & 	      &	     \\
\hline                    
\end{tabular}
\tablefoot{$RV_{\mathrm{P}}$, $RV_{\mathrm{S}}$, and $RV_{\mathrm{T}}$ are the radial velocities measured by cross-correlation, expressed in the heliocentric reference frame (no correction for the systemic velocities has been applied). We stress that the error bars on these values are about 5~\kms.}
\end{table*}

Equivalent widths (EWs) and the Doppler shifts were estimated on each spectrum by fitting Gaussian profiles on the spectral lines. For the computation of the radial velocities, we adopted the rest wavelengths from \citet{con77}, for the lines with a rest wavelength shorter than 5000~\AA, and \citet{und94}, for the other ones. 


\section{Third component and disentangling}
\label{s_dis}
\begin{figure}[htbp]
\centering
\includegraphics[width=8cm,bb=20 158 585 710,clip]{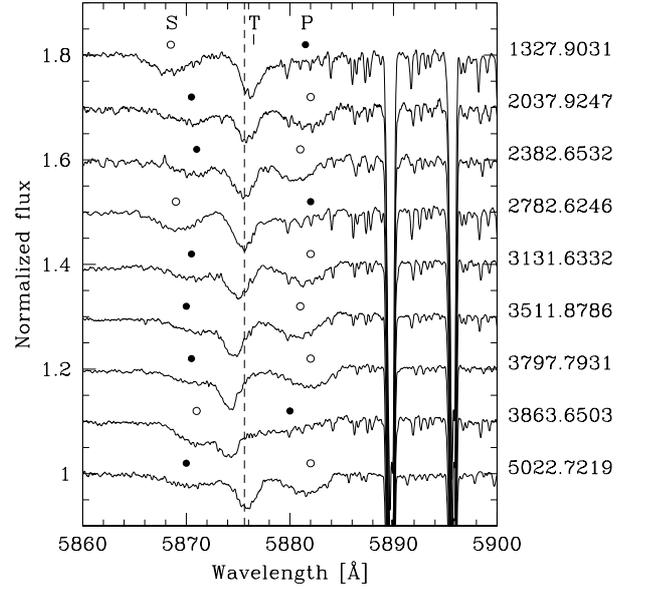}
\caption{Normalized spectra of HD\,150136 zoomed on the [$5860-5900$]~\AA\ region and vertically shifted for clarity. The \ion{He}{i}~$\lambda{5876}$ line displays a clear triple signature. Primary (P) and secondary (S) components are represented by filled and open circles, respectively.}\label{fig_spec}
\centering
\end{figure}

On the basis of the high-resolution spectra analysed in the present paper, we report the unambiguous detection of a third star (T) signature for the system HD\,150136 (Fig.~\ref{fig_spec}). The data indeed exhibit a clear variation in the third component compared to the rest wavelength (\ion{He}{i}~$\lambda{5875.62}$) shown by the dashed line. Moreover, the last observation confirms the periodicity of the variation, suggesting that the third star is also a member of a binary system. Although several analyses have suspected the existence of this component, it has never been unveiled until now.

\begin{figure*}[htbp]
\centering
\includegraphics[width=15cm,bb=38 172 576 698,clip]{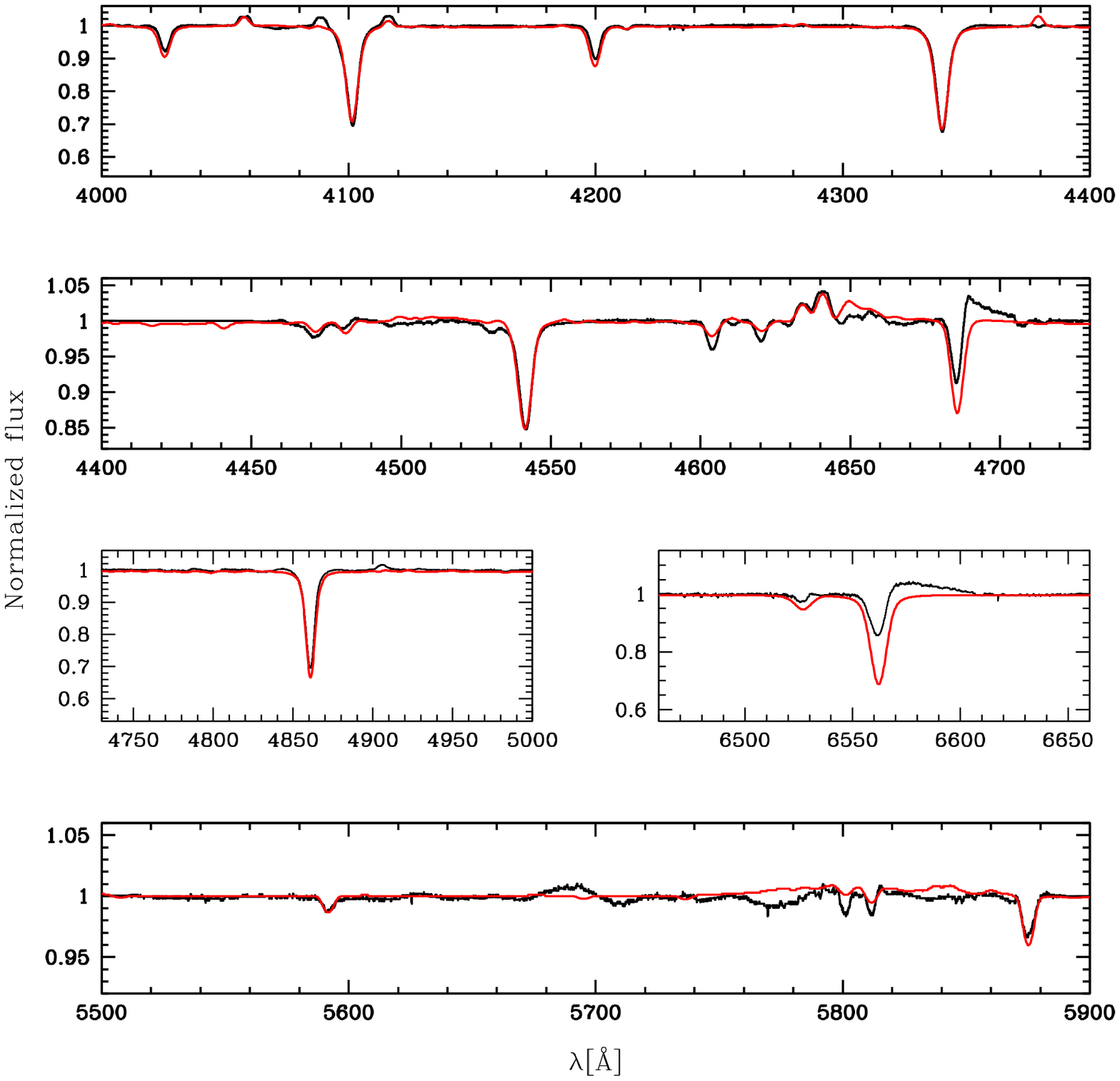}
\caption{Best-fit CMFGEN models (red) of the disentangled optical spectrum (black) of the primary component of HD\,150136.}\label{fit_op_prim}
\centering
\end{figure*}

 To separate the contribution of each star in the observed spectra, we apply a modified version of the disentangling program of \citet{gl06} by adapting it to triple systems. This method consists of an iterative procedure that alternatively uses the spectrum of a first component (shifted by its radial velocity) and then the spectrum of a second component (also shifted by its radial velocity) to successively remove them from the observed spectra. The obtained result is a mean spectrum of the remaining third star. This program also allows us to recompute the radial velocities of each star with a cross-correlation technique, even at phases for which the lines are heavily blended. The cross-correlation windows are built on a common basis that gathers the positions of the \ion{He}{ii}~$\lambda{4200}$, \ion{He}{i}~$\lambda{4471}$, \ion{He}{ii}~$\lambda{4542}$, \ion{He}{i}~$\lambda{4713}$, \ion{He}{i}~$\lambda{4921}$, \ion{O}{iii}~$\lambda{5592}$, \ion{C}{iv}~$\lambda\lambda{5801-12}$, and \ion{He}{i}~$\lambda{5876}$ spectral lines. In addition, we add the \ion{N}{v}~$\lambda\lambda{4603-19}$ lines for the primary star. The radial velocities obtained with this technique are listed in Table~\ref{tab1}. The resulting spectra are disentangled in a [$4000-5900$]~\AA\ wavelength domain and at H$\alpha$. Though the \ion{He}{ii}~$\lambda{4686}$ and H$\alpha$ lines present particular variations along the orbital phase, we include them in the process. However, the assumption of constant lines (on which relies the disentangling method) is no longer satisfied, thus making these line profiles unreliable.

\begin{figure*}[htbp]
\centering
\includegraphics[width=15cm,bb=38 172 576 698,clip]{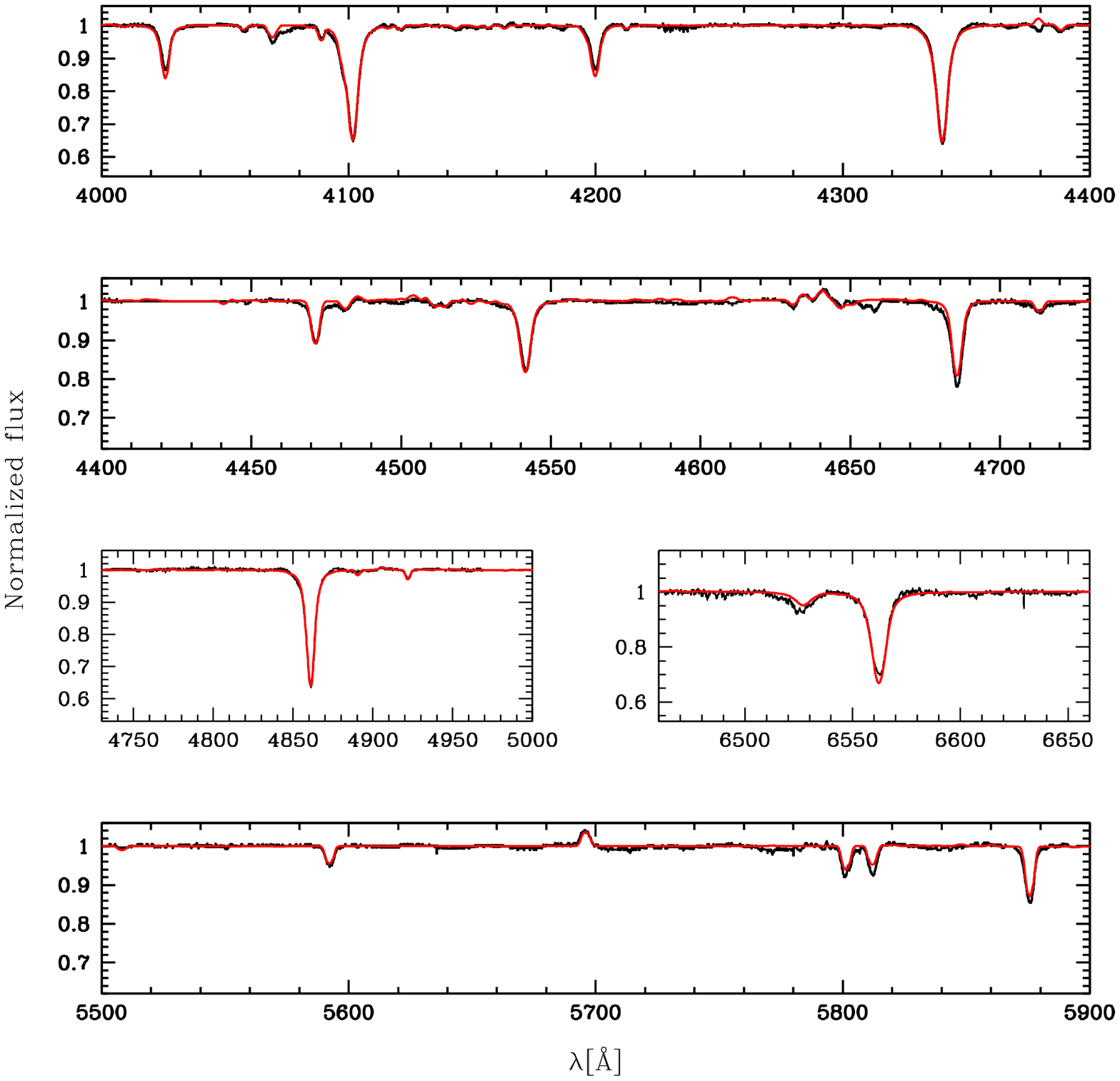}
\caption{Same as for Fig.~\ref{fit_op_prim} but for the secondary component of HD\,150136.}\label{fit_op_seco}
\centering
\end{figure*}

\begin{figure*}[htbp]
\centering
\includegraphics[width=15cm,bb=38 172 576 698,clip]{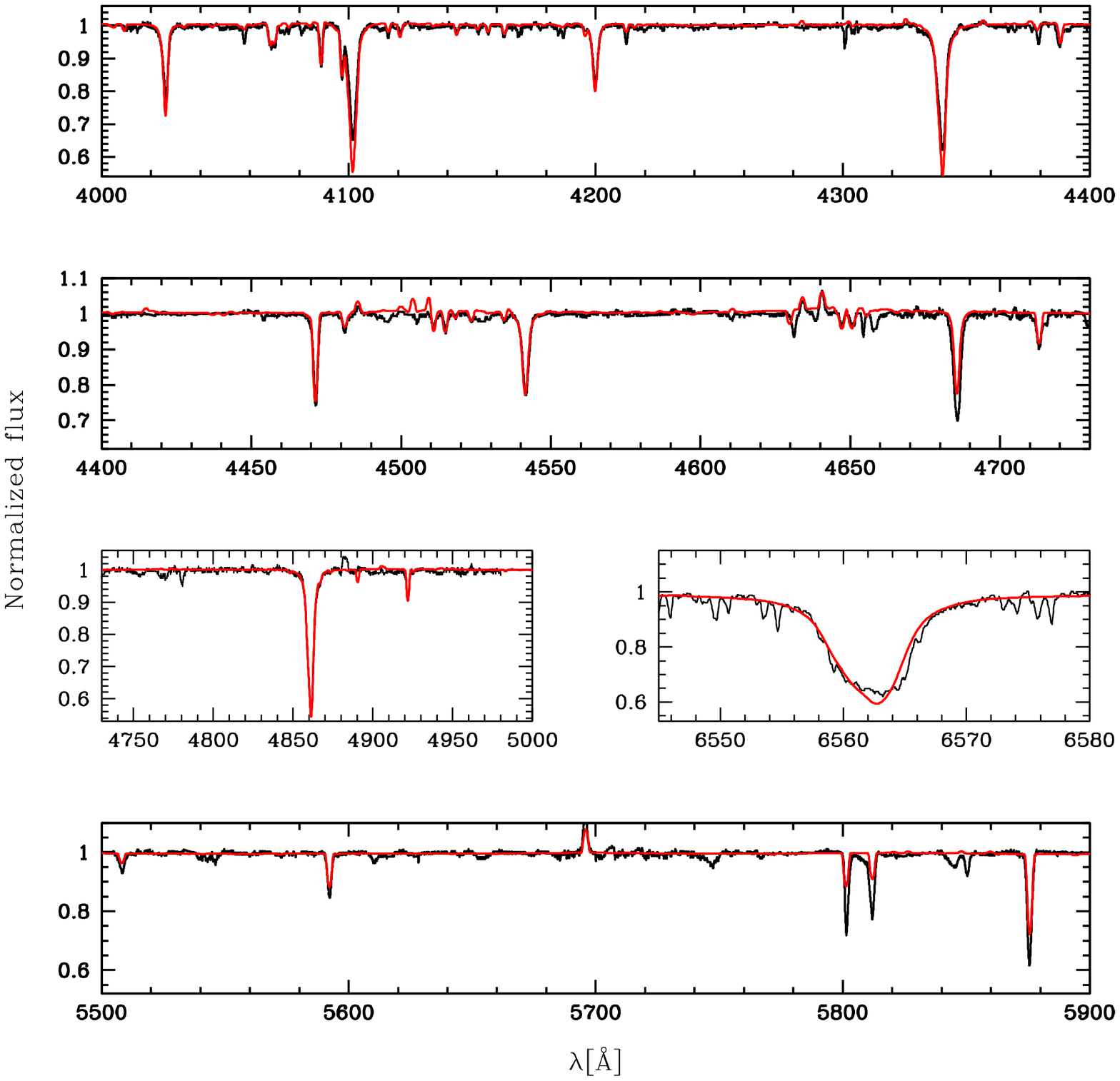}
\caption{Same as for Fig.~\ref{fit_op_prim} but for the third component of HD\,150136.}\label{fit_op_thir}
\centering
\end{figure*}

To obtain the final individual spectra, we have to correct the disentangled spectra by the brightness ratio. For that purpose, we use the disentangled and the observed spectra to measure the EWs and constrain the spectral classification of the three components of HD\,150136. This determination is based on the quantitative classification criteria presented by \citet{con71} and \citet{con73b}, which uses the EW ratios of the \ion{He}{i}~$\lambda{4471}$ and \ion{He}{ii}~$\lambda{4542}$ lines to define the spectral type. This leads to an O3--O3.5 primary star, an O5.5--O6 secondary star and an O6.5--O7 third star. The weak \ion{N}{iii}~$\lambda\lambda{4634-41}$ emission and the strong \ion{He}{ii}~$\lambda{4686}$ absorption lines of the secondary and the tertiary components suggest adding the ((f)) suffix to their spectral types. In addition, the criterion given by \citet{wal02} confirms the classification of the primary. Indeed, its spectrum is reminiscent of that of HD\,64568, i.e., close to an O3V((f$^{*}$)). However, the \ion{N}{iv}~$\lambda{4058}$ line has an intermediate EW ($\sim 160$~m\AA) between the values measured on the \ion{N}{iii}~$\lambda{4634}$ ($\sim 140$~m\AA) and \ion{N}{iii}~$\lambda{4641}$ ($\sim 250$~m\AA) lines, thereby indicating a spectral type between O3V((f$^{*}$)) and O3.5V((f$^{+}$)). The ((f$^{*}$)) reports a spectrum with the \ion{N}{iv}~$\lambda{4058}$ emission line stronger than the \ion{N}{iii}~$\lambda\lambda{4634-41}$ lines and a weak \ion{He}{ii}~$\lambda{4686}$ absorption line whilst the ((f$^{+}$)) refers to a spectrum with medium \ion{N}{iii}~$\lambda\lambda{4634-41}$ emission lines, the weak \ion{He}{ii}~$\lambda{4686}$ line, and the \ion{Si}{iv}~$\lambda\lambda{4089-4116}$ lines in emission. To compute the different brightness ratios required to determine the relative luminosities of the stars, we compare the EWs measured on the observed spectra with the theoretical values issued from synthetic models whose spectral classification is the same as for our disentangled spectra. We stress that these models are computed from solar surface abundances (except for the nitrogen abundance). To avoid a possible bias due to differences between fits and models, we also focus on tables of \citet{con71}, \citet{con73}, and \citet{con74} for the secondary and the tertiary components (the spectral type of the primary is not available in these tables). This comparison between the ``observed'' EWs and the ``canonical'' EWs for isolated single stars is performed on the helium and oxygen lines, yielding brightness ratios of $l_{\mathrm{P}}/l_{\mathrm{S}}=2.54$, $l_{\mathrm{S}}/l_{\mathrm{T}}=1.51$, and $l_{\mathrm{P}}/l_{\mathrm{T}}=3.87$. An error bar at 1~-~$\sigma$ of about 0.14 is estimated between the ``observed/canonical'' ratios measured from the different spectral lines. The individual spectra, corrected from the brightness ratios, are shown as black lines in Figs.~\ref{fit_op_prim}~to~\ref{fit_op_thir}.

\section{Inner short-period binary system}
\label{s_inner}
\begin{figure}[htbp]
\centering
\includegraphics[width=9cm,bb=-140 151 720 618,clip]{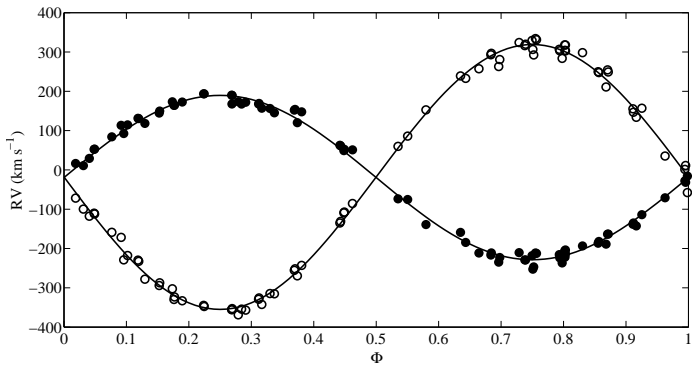}
\caption{Radial-velocity curves corresponding to the short-period inner SB2 system as a function of the phase. Filled circles (open) represent the primary (secondary) component (see Sect.~\ref{s_inner} for details).}\label{so_1}
\centering
\end{figure}

With the radial velocities refined by the disentangling program, we compute the orbital solution of the short-period system formed by the primary (P) and secondary (S) components. We first re-estimate the orbital period of the short-term binary by using the Heck-Manfroid-Mersch technique (hereafter HMM, \citealt{hec85}, as revised by \citealt{gos01}) on the basis of $RV_{\mathrm{S}}-RV_{\mathrm{P}}$ values. We obtain an orbital period of about $2.67456\pm0.00001$. We then use the Li{\`e}ge Orbital Solution Package (LOSP\footnote{LOSP is developed and maintained by H. Sana and is available at http://www.science.uva.nl/$\sim$hsana/losp.html. The algorithm is based on the generalization of the SB1 method of \citet{wol67} to the SB2 case along the lines described in \citet{rau00} and \citet{san06a}.}) to determine the SB2 orbital solution of the inner binary. Figure~\ref{so_1} displays the radial-velocity curves, and Table~\ref{tab_orb} lists the orbital parameters determined for an SB2 configuration. The system shows no significant deviation (test of \citealt{luc71}) from circularity, so we thus adopt $e=0$. As a consequence, $T_0$ represents the phase of the conjunction with the primary in front. We see that the dispersion of some observational points around the theoretical radial-velocity curves is rather large, which is probably due to the influence of the third star.

\begin{figure}[htbp]
\centering
\includegraphics[width=7cm,bb=22 13 524 394,clip]{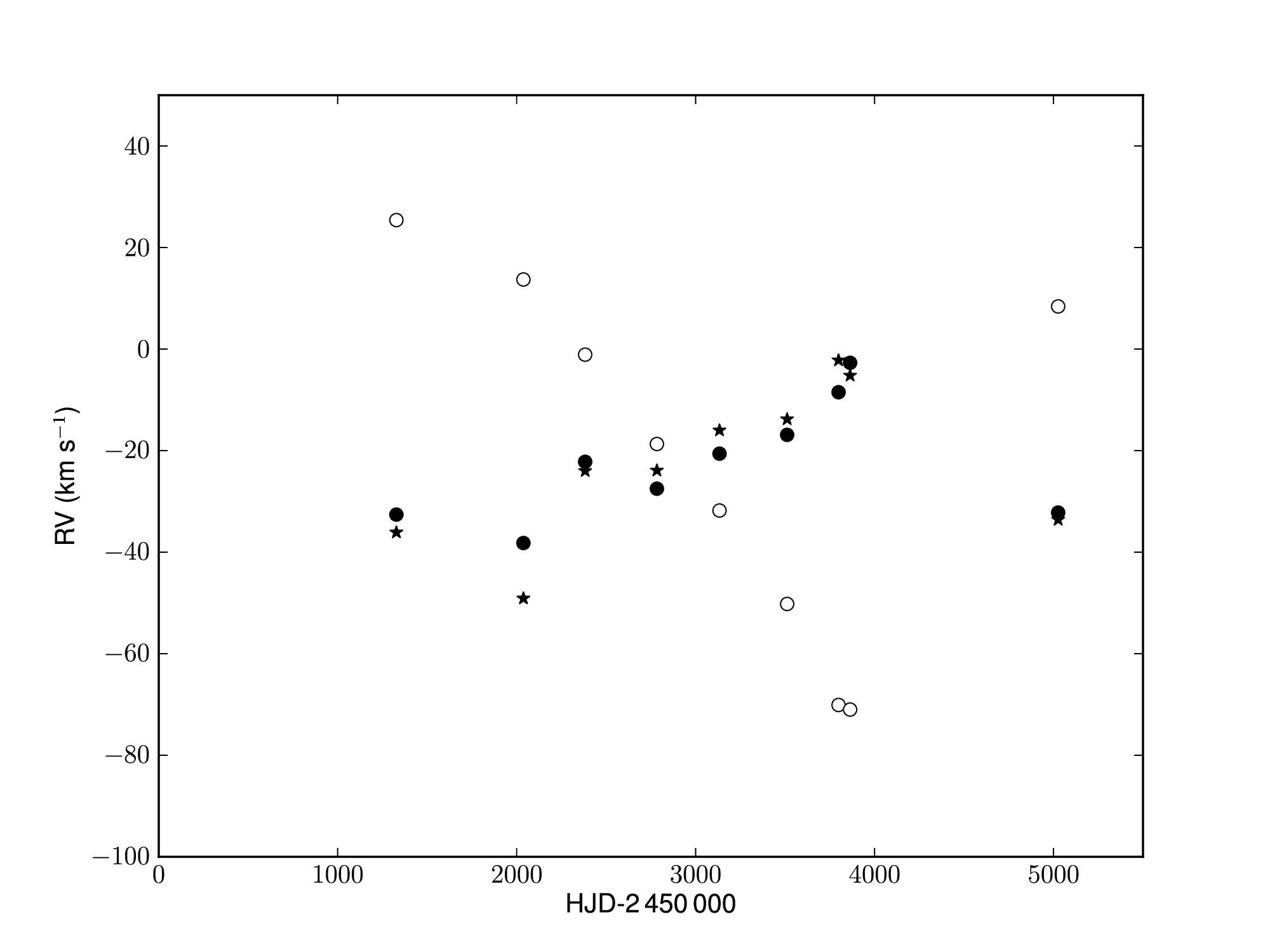}
\caption{Radial velocities of the outer system represented as a function of the time. Filled circles/filled stars correspond to the systemic velocities of the primary/secondary component, whilst the open circles report the mean radial velocities of the third component (more details are provided in Sect.~\ref{s_inner}).}\label{so_2}
\centering
\end{figure}

Consequently, we compute the averages, for the primary and the secondary stars over each epoch, of the differences between the observations and the fitted radial-velocity curves. We also add to these values the constant systemic velocities quoted in Table~\ref{tab_orb}. Moreover, we compare these results to the systemic velocities of the primary and the secondary components computed on each individual observing campaign. Both methods report comparable global trends of the velocities for the inner system. The adopted values are reported in Table~\ref{vrad_third}. By plotting the mean systemic velocities of the inner system (see Table~\ref{vrad_third}) and the mean radial velocities of the third component computed on each epoch as a function of time (Fig.~\ref{so_2}), we detect a clear trend toward a periodical motion which is reminiscent of an additional orbit. Therefore, we suggest that HD\,150136 is at least an SB3 system. 

\begin{table*}
\begin{center}
\caption{Orbital solution for the short-period binary system computed with LOSP and an SB3 program as described in Sects.~\ref{s_inner} and~\ref{s_outer}.} \label{tab_orb}
\begin{tabular}{lrrrr}
\hline\hline
                    & \multicolumn{2}{c}{SB2 solution}& \multicolumn{2}{c}{SB3 solution}\\
                    & Primary        & Secondary      & Primary        & Secondary \\
\hline
$P$ [day] & \multicolumn{2}{c}{2.67454 $\pm$ 0.00001} & \multicolumn{2}{c}{2.67455 $\pm$ 0.00001}\\
$e$ &   \multicolumn{2}{c}{0.0 (fixed)}  &   \multicolumn{2}{c}{0.0 (fixed)}\\
$T_0$ [HJD~$-$~2\,450\,000]  & \multicolumn{2}{c}{1327.164 $\pm$ 0.003}  & \multicolumn{2}{c}{1327.161 $\pm$ 0.006}\\
$q (M_1/M_2)$ & \multicolumn{2}{c}{1.615 $\pm$ 0.021}  & \multicolumn{2}{c}{1.586 $\pm$ 0.019}\\
$\gamma$ [\kms] & $-$19.7 $\pm$ 1.5 & $-$17.8 $\pm$ 1.8  & $-$26.1 $\pm$ 2.5 & $-$24.4 $\pm$ 2.5\\
$K$ [\kms] &  208.9 $\pm$ 1.7 & 337.2 $\pm$ 2.7  &  211.2 $\pm$ 1.5 & 335.0 $\pm$ 1.5\\
$a \sin i$ [$R_{\odot}$] & 11.0 $\pm$ 0.1 & 17.8 $\pm$ 0.1 & 11.2 $\pm$ 0.1 & 17.7 $\pm$ 0.1\\
$M \sin^3 i$ [\msun] & 27.9 $\pm$ 0.5 & 17.3 $\pm$ 0.3 & 27.7 $\pm$ 0.4 & 17.5 $\pm$ 0.3\\
rms [\kms] & \multicolumn{2}{c}{$13.15$} & \multicolumn{2}{c}{$7.95$}\\
\hline
\end{tabular}
\tablefoot{The given errors correspond to 1~-~$\sigma$.}
\end{center}
\end{table*}

\begin{table*}
\begin{center}
\caption{Systemic velocities of the primary and secondary components, as well as the mean value over each epoch (for more explanations see Sect.~\ref{s_inner}).} \label{vrad_third}
\begin{tabular}{lcrcrcr}
\hline\hline
HJD              & \# obs & Primary & Error Primary & Secondary & Error Secondary & Mean\\
$- 2\,450\,000 $ &        &  [\kms] &  [\kms]       &  [\kms]   &  [\kms]         & [\kms]\\
\hline
1327.9031 & 1 & $-32.6$ & --  & $-36.1$ & --  & $-34.4$\\
2037.8496 & 2 & $-38.2$ & 12.5& $-49.1$ & 27.0& $-43.7$\\
2382.7319 & 6 & $-22.2$ &  4.6& $-24.0$ &  5.5& $-23.1$\\
2783.6601 & 3 & $-27.5$ &  2.2& $-23.9$ &  2.5& $-25.7$\\
3133.4644 & 21& $-20.6$ &  1.4& $-16.0$ &  1.7& $-18.3$\\
3511.4421 & 11& $-16.9$ &  0.8& $-13.8$ &  1.2& $-15.4$\\
3798.7490 & 13& $ -8.5$ &  1.2&  $-2.2$ &  1.5& $ -5.4$\\
3863.3497 & 7 & $ -2.7$ &  1.3& $ -5.2$ & 1.7 & $ -3.9$\\
5025.8661 & 15& $-32.2$ &  1.2& $-33.6$ & 1.8 & $-32.9$\\
\hline
\end{tabular}
\tablefoot{The quoted errors correspond to 1~-~$\sigma$. Second column represents the number of observations per epoch and the last five columns give the systemic velocity of the primary, the error on this value, the systemic velocity of the secondary, the error on this value, and the average on the primary and secondary systemic velocities, respectively.}
\end{center}
\end{table*} 

\section{The outer system}\label{s_outer}
\subsection{The outer system as an SB2}

We use the mean systemic velocities of the inner system determined from each epoch as a primary component, and we consider the third component as the secondary companion. These velocities do not allow us to determine the real orbital period of the wide outer system with accuracy. However, we can estimate this value between roughly 2950 and 5500~d, i.e, between 8 and 15 years. Consequently, computing an orbital solution is useless without further high-resolution monitoring of HD\,150136. Nonetheless, the measured velocities can provide information about the orbit, as well as on the mass ratio between the inner system and the third component. To estimate the mass ratio, we calculate the average between the systemic velocities of the primary and the secondary stars and we take these values as radial velocities of the inner system (P$+$S). We focus on the radial velocities of both outer system components by fitting the following relation by a least-square method: 
\begin{equation*}
RV_{\mathrm{T}}(\phi) = c_1\,RV_{\mathrm{P+S}}(\phi)+c_2
\end{equation*}
where $RV_{\mathrm{P+S}}(\phi)$, $RV_{\mathrm{T}}(\phi)$ are the systemic radial velocities of the inner system and the radial velocities of the third component, respectively, whereas $c_1 = -\frac{M_{\mathrm{P+S}}}{M_{\mathrm{T}}}$ with $M_{\mathrm{P+S}}$ and $M_{\mathrm{T}}$ the masses of the inner system and the third star, respectively. The linear regression gives us values of about $c_1 = -2.60$ and $c_2 = -79.58$~\kms. From $c_2$, we deduce $\gamma_{\mathrm{T}} \approx -20$~\kms. To verify the impact of the period on the value of the mass ratio, we have also computed several orbital solutions by assuming different orbital periods sampled between 2950 and 5500~d. All these solutions led to a mass ratio of about 2.94. However, among the velocities exhibited in Fig.~\ref{so_2}, the point corresponding to HJD~=~2\,452\,037.8496 (i.e., to the observing campaign made in 2001) displays a clear deviation probably due to the shortage of observations during this campaign. If we remove this point, we obtain by the linear-regression method a mass ratio of about 3.1 and by the other method a mass ratio of about 3.2. We thus conclude on a mass ratio of about $2.9\pm0.3$ between the inner system and the third star. Similarly, from this sample of orbital periods, we derive an eccentricity higher than 0.3 for the outer system.

\subsection{HD\,150136 as a triple system}

\begin{figure}[htbp]
\centering
\includegraphics[width=9cm,bb=-140 151 720 618,clip]{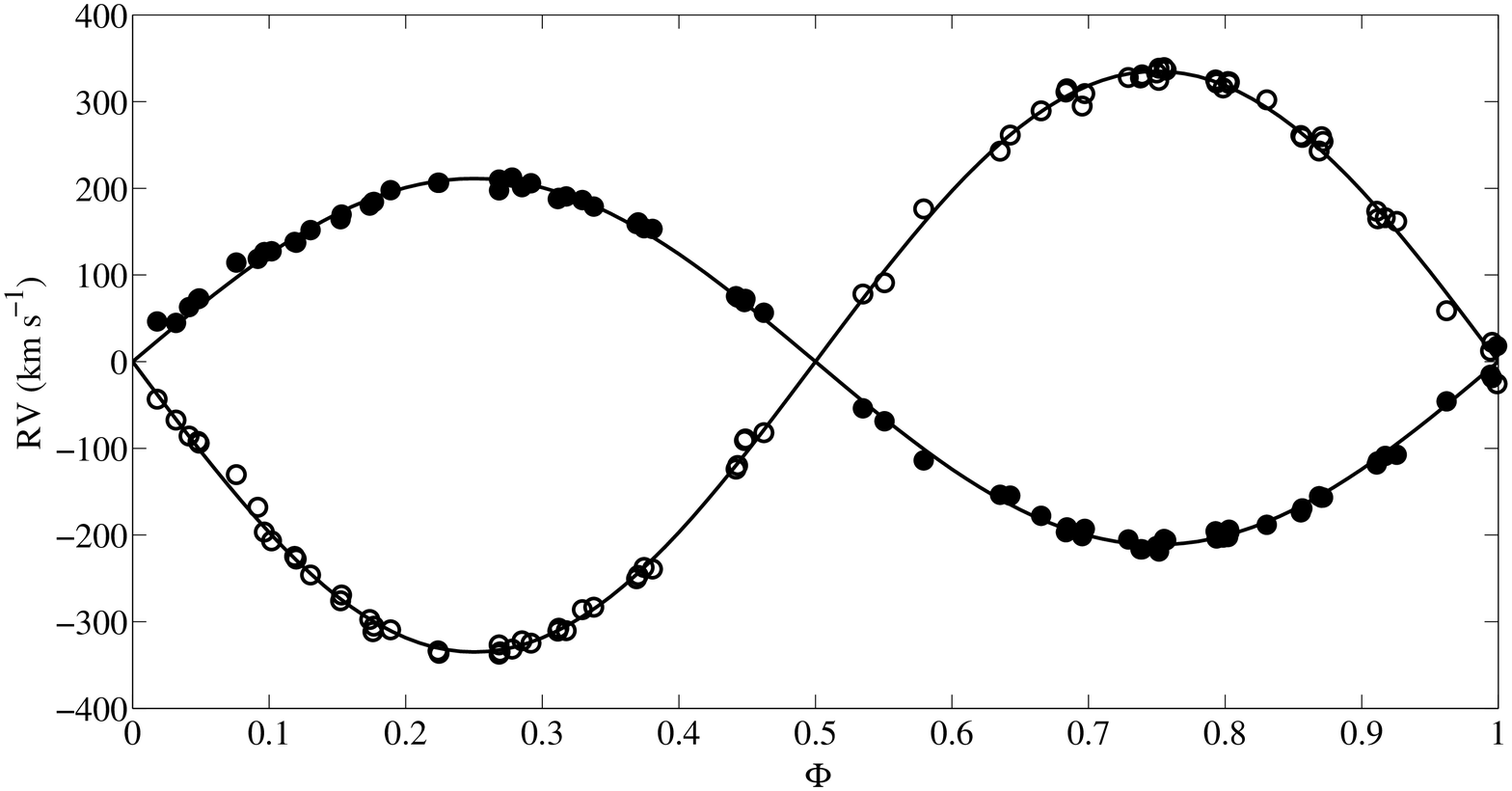}
\caption{Radial-velocity curves corresponding to the short-period inner system as a function of the phase according to the SB3 fit. Filled circles (open) represent the primary (secondary) component. The velocities are expressed in the frame of the centre of mass of the P$+$S system (see Sect.~\ref{s_outer}).}\label{so_1_refined}
\centering
\end{figure}

To better constrain the orbital parameters of the inner system, we take the influence of the third component into account by fitting all the components together. For that purpose, we determine with another method the orbital parameters of the inner system. This technique provides the approximate orbital solutions for the two SB2 systems in HD\,150136, i.e., the inner system composed of the primary and the secondary stars and the outer one represented by the mean systemic velocities characterizing the centre of mass of the inner system and the radial velocities of the third component. Therefore, we perform on all the three sets of data (P, S, and T) a global least square fit with an assumed circular orbit for the inner binary. For both systems, we follow the same procedure. We first deduce the approached orbital solution using the keplerian periodogram introduced by \cite{zec09}. This method consists in a fit of a simple sine function in the true anomaly domain. The approximate orbital solution is then globally refined using a Levenberg-Marquardt minimization. We search the period in [0.6$T$,2.0$T$] for the outer orbit where $T$ is the total time span of the observations. We use a constant step of frequency for both systems equal to $0.01 / T$.
Figure~\ref{so_1_refined} exhibits the orbital solution for the short-term binary system by considering the influence of the third star. We report in Table~\ref{tab_orb} the refined orbital parameters (SB3 solution). Among these values, we notably observe a decrease in the error bars. Following an F-test on the ratio of the two $\chi^2$ of the fits (SB2 and SB3 solutions, see Table~\ref{tab_orb}), we conclude that this solution is clearly improved in comparison with the one computed in Sect.~\ref{s_inner} (see Fig.~\ref{so_1}). The improvement is sufficiently marked to remain significant even though the F-test is not fully valid due to the non-linearity. 

\section{Atmosphere analysis of the three stars}
\label{s_spec}

The three components of HD\,150136 are then modelled separately by using the CMFGEN atmosphere code \citep{hm98}. Non-LTE models including winds and line-blanketing are computed. The computations are done in spherical geometry and in the co-moving frame. A more detailed description of this code is provided in \citet{hm98}. For the velocity structure, we use TLUSTY hydrostatic structures in the photosphere associated with a $\beta$-velocity law $v = v_{\infty} (1-R/r)^{\beta}$ for the wind part (\vinf\ being the wind terminal velocity). The TLUSTY models are taken from the OSTAR2002 grid of models computed by \citet{lh03} and $\beta = 0.8$, typical for O dwarfs, is adopted \citep{rep04}. The models include the following chemical species: H, He, C, N, O, Ne, Mg, Si, S, Ar, Fe, and Ni with the solar composition of \citet{gas07}, except for N where the abundance is kept free. This is also the case for C and O abundances, but we see below that they are not well constrained and so are fixed to the solar values. We also use the superlevel approach to limit the required computation memory. When the model atmosphere has converged, the synthetic spectrum is generated through the formal solution of the radiative transfer equation. For that purpose, a microturbulent velocity varying linearly (with velocity) from 10 \kms, in the photosphere, to 0.1 $\times$ \vinf, in the outer boundary, is used. The emerging spectra are compared to the disentangled ones, corrected for the brightness ratios, in order to constrain the main parameters more accurately.

We thus determine the effective temperature (\teff), the $\log g$, the luminosity, the N content, the projected rotational velocity, and the macroturbulence of each component with the following diagnostics:
\begin{itemize}
\item the {\it effective temperature}: \teff\ is computed on the basis of the ratio between the \ion{He}{i}~$\lambda{4471}$ and \ion{He}{ii}~$\lambda{4542}$ lines. However, other lines of helium such as \ion{He}{i}~$\lambda{4026}$, \ion{He}{ii}~$\lambda{4200}$, \ion{He}{i}~$\lambda{4388}$, \ion{He}{i}~$\lambda{4713}$, \ion{He}{i}~$\lambda{4921}$, and \ion{He}{i}~$\lambda{5876}$ are also considered as indicators. Generally, we estimate the error in the models on the determination of that parameter to be equal to 1000~K.\\
\item the {\it gravity}: we use the wings of the Balmer lines (H$\delta$, H$\gamma$, and H$\beta$) as main diagnostics to constrain the $\log g$. We refrain from using the H$\alpha$ line because it forms, at least partially, in the stellar wind. For the tertiary star, the determination of the gravity is more uncertain because of the disentangling process (see Sect.~\ref{s_disc}). Therefore, the uncertainty on the \logg\ is equal to 0.1 dex for the primary and the secondary and to 0.25 dex for the third component.\\
\item the {\it luminosity}: the luminosity is constrained by the V-band magnitude by adopting a distance of 1.32~kpc and a colour excess $E(B-V)$ of $0.44$ as explained in Sect.~\ref{s_dis}. The accuracy on the luminosity depends on the uncertainty on the distance of HD\,150136.\\
\item the {\it surface abundances}: the nitrogen abundance is determined on the \ion{N}{iii} triplet in the $4500-4520$~\AA\ range but also on the \ion{N}{iv}~$\lambda{4058}$ and \ion{N}{iii}~$\lambda\lambda{4634-41}$ line profiles. The typical uncertainty on the CNO surface abundances is of about 50\%. To estimate these uncertainties, we run models with several different values of N and compare the corresponding spectra to the observed line profiles. When a clear discrepancy is seen on all lines of the same element, we adopt the corresponding abundance as the maximum/minimum value of the possible chemical composition. Carbon abundance is estimated to be solar for the three stars, whilst oxygen abundance is not determined because of the lack of diagnostic lines. \\
\item the {\it rotational and macroturbulent velocity}: the synthetic spectra are successively convolved by a rotational profile and then by a Gaussian profile representing the macroturbulence. First, the projected rotational velocity is computed with the method of \citet{sim07} based on the Fourier transform. Then, the macroturbulent velocity is determined so that the shape of the \ion{He}{i}~$\lambda{4713}$ line is reproduced best. When this line is not visible in the spectrum, we use the \ion{He}{i}~$\lambda{5876}$ and/or the \ion{C}{iv} doublet $\lambda \lambda 5801-12$ as second indicator.
\end{itemize}

From $V=5.5\pm0.12$ \citep{ng05}, a mean $(B-V)=0.16\pm0.01$, measured from Reed's catalogue \citep{ree05}, and a $(B-V)_0 = -0.28$ \citep{mar06}, we compute an absolute $V$ magnitude for the entire system equal to $M_V = -6.47_{-0.36}^{+0.34}$ by assuming a distance of about $1.32\pm0.12$~kpc. From the different brightness ratios, we then obtain $M_{V_{\mathrm{P}}} = -5.91_{-0.36}^{+0.35}$, $M_{V_{\mathrm{S}}} = -4.90_{-0.36}^{+0.35}$ and $M_{V_{\mathrm{T}}} = -4.44_{-0.37}^{+0.36}$ for the primary, the secondary and the tertiary stars, respectively. The bolometric correction is computed on the basis of the expression given by \citet{mar05} from a typical effective temperature determined as a function of the spectral type of the three stars (see Table~\ref{tab_mod}). These values coupled with the absolute $V$ magnitude of the three stars yield the bolometric luminosities of about $\log \left( L_{\mathrm{P}}/L_{\odot}\right) = 5.86_{-0.11}^{+0.12}$, $\log \left( L_{\mathrm{S}}/L_{\odot}\right) = 5.32_{-0.10}^{+0.12}$, and $\log \left( L_{\mathrm{T}}/L_{\odot}\right) = 5.01_{-0.10}^{+0.11}$ for the three stars. These luminosities are in good agreement with the ones of main-sequence stars having similar spectral classifications.

The best-fit model spectra of the three components are displayed in red in Figs.~\ref{fit_op_prim} to~\ref{fit_op_thir}, whilst the derived parameters are given in Table~\ref{tab_mod}. However, the optical spectra do not give us enough information to constrain the wind properties of these components. We accordingly limit our investigation on the stellar parameters. Indeed, the H$\alpha$ and \ion{He}{ii}~$\lambda{4686}$ lines are likely to be affected (to some extent) by the colliding-wind interaction zone, at least for the primary and the secondary components. Since the disentangling generates mean spectra on the entire dataset, the resulting H$\alpha$ and \ion{He}{ii}~$\lambda{4686}$ line profiles could therefore be different from those of single stars with the same spectral classifications. We stress that, as we see in Fig.~\ref{fit_op_prim}, both lines display P\,Cygni profiles for the primary, which means that the stellar wind of this star could be rather strong in comparison to its spectral classification. Therefore, we also investigate the UV domain, but the small number and the poor resolution of IUE data in the archives do not allow us to separate the individual contributions of the three components. Given the large ionizing flux generated by an O3 star, we thus assume that the flux collected in UV is mainly from the primary, whilst the \ion{He}{ii}~$\lambda{4686}$ and H$\alpha$ lines in the optical range serve to constrain the wind parameters of the secondary and the third components. Therefore, we recommend considering the wind parameter values (\mdot\ and \vinf) listed in Table~\ref{tab_mod} as preliminary ones useful for computing the model atmospheres.

\begin{table}
\caption{Stellar parameters of the three components.}
\label{tab_mod} 
\centering      
\begin{tabular}{lrrr}
\hline\hline         
Parameters & Primary & Secondary & Tertiary \\  
\hline                                   
Sp. type        & O3V((f$^{*}$))/   & O5.5--6V((f))       & O6.5--7V((f)) \\
& O3.5V((f$^{+}$)) & & \\
\teff\ [K]      & $46500\pm1000$ & $40000\pm1000$ & $36000\pm1000$\\      
$\log \left( \frac{L}{L_{\odot}}\right)$ & $5.86_{-0.11}^{+0.12}$ & $5.32_{-0.10}^{+0.12}$ & $5.01_{-0.10}^{+0.11}$  \\[1pt]
$R$\ [$\rsun$]           &  $13.14_{-2.05}^{+3.62}$ & $9.54_{-1.45}^{+1.98}$ & $8.24_{-1.29}^{+1.65}$\\[1pt]
$\log g$ [cgs]           & $4.00\pm0.10$  & $4.00\pm0.10$  & $3.50\pm0.25$\\[1pt]
$M_{\mathrm{spec}}$\ [$\msun$]        & $63.9_{-27.4}^{+27.5}$ & $33.2_{-14.2}^{+27.7}$ & $7.8_{-4.7}^{+12.3}$ \\[1pt]
$M_{\mathrm{evol}}$\ [$\msun$]        & $70.6_{-9.1}^{+11.4}$ & $36.2_{-1.6}^{+5.0}$ & $27.0_{-3.5}^{+3.0}$ \\[1pt]
$M\,\sin^3{i}$ [$\msun$]    &   $27.7 \pm 0.4$ & $17.5 \pm 0.3$    &       --            \\[1pt]
N/H\ [$\times 10^{-4}$]    & 8.0$\pm$4.0 &  3.0$\pm$1.5  & 3.0$\pm$2.0 \\[1pt]
\vsini\ [\kms]            &  171$\pm$20  &  136$\pm$20  & 72$\pm$10\\[1pt]
$v_{\mathrm{mac}}$\ [\kms]          &  65$\pm$10   &  41$\pm$10   & 14$\pm$5 \\[1pt]
$\dot{M}$ [$\msun$ yr$^{-1}$]&  $1.0\times10^{-6}$   &  $7.0\times10^{-8}$ &  $3.0\times10^{-9}$\\[1pt]
$v_{\infty}$\ [\kms]       &  $3500\pm100$     &  $2500\pm100$      & $1200\pm100$\\[1pt]
\hline                                             
\end{tabular}
\tablefoot{The given errors correspond to 1~-~$\sigma$. The solar abundance for nitrogen is $\mathrm{N/H}=0.6\times 10^{-4}$. }
\end{table}


\section{Doppler tomography of the inner system}
\label{s_tomo}

\begin{figure*}[htbp]
\centering
\includegraphics[width=8cm,bb=30 173 585 696,clip]{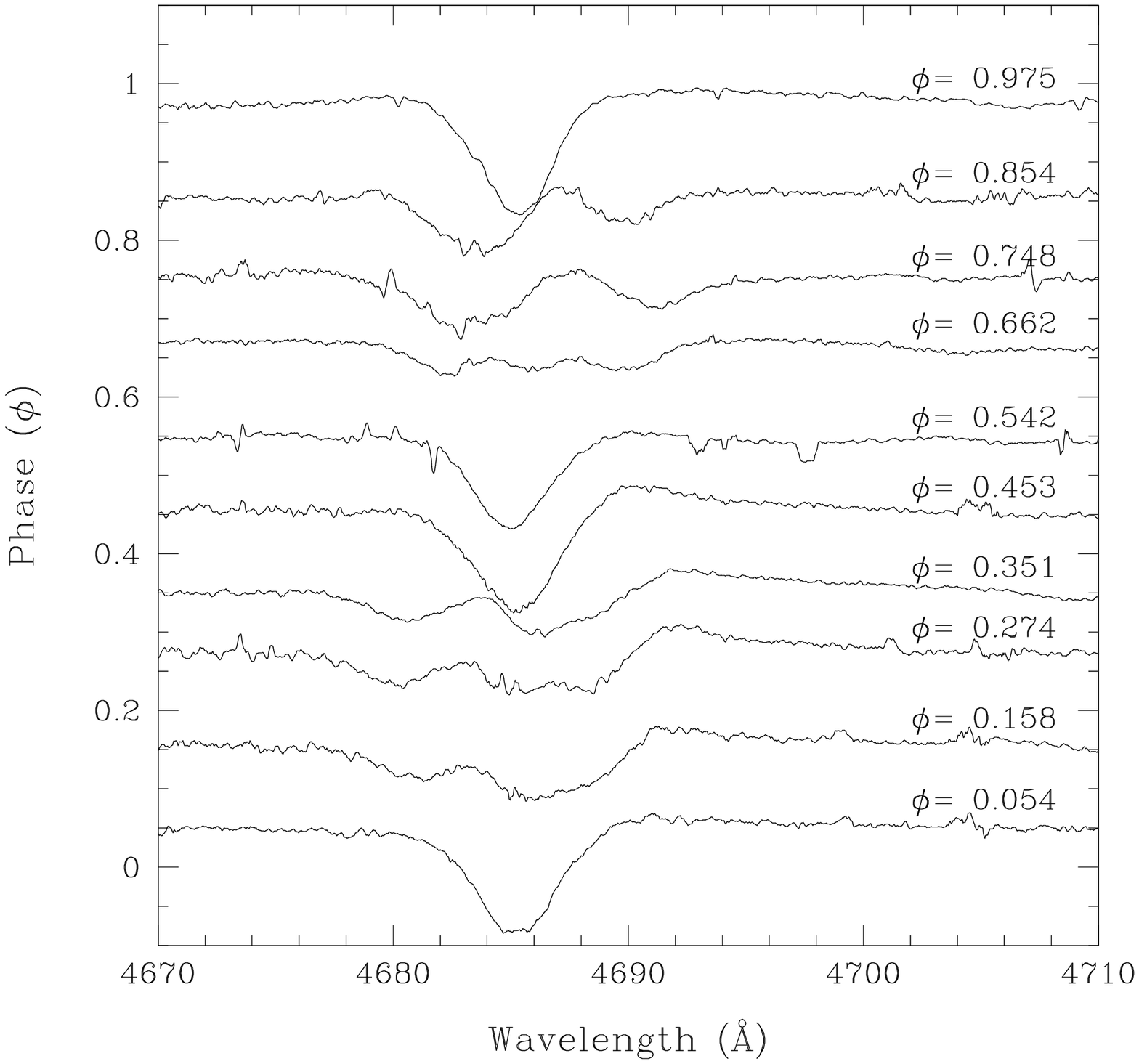}
\hspace{2cm}\includegraphics[width=8cm,bb=30 173 585 696,clip]{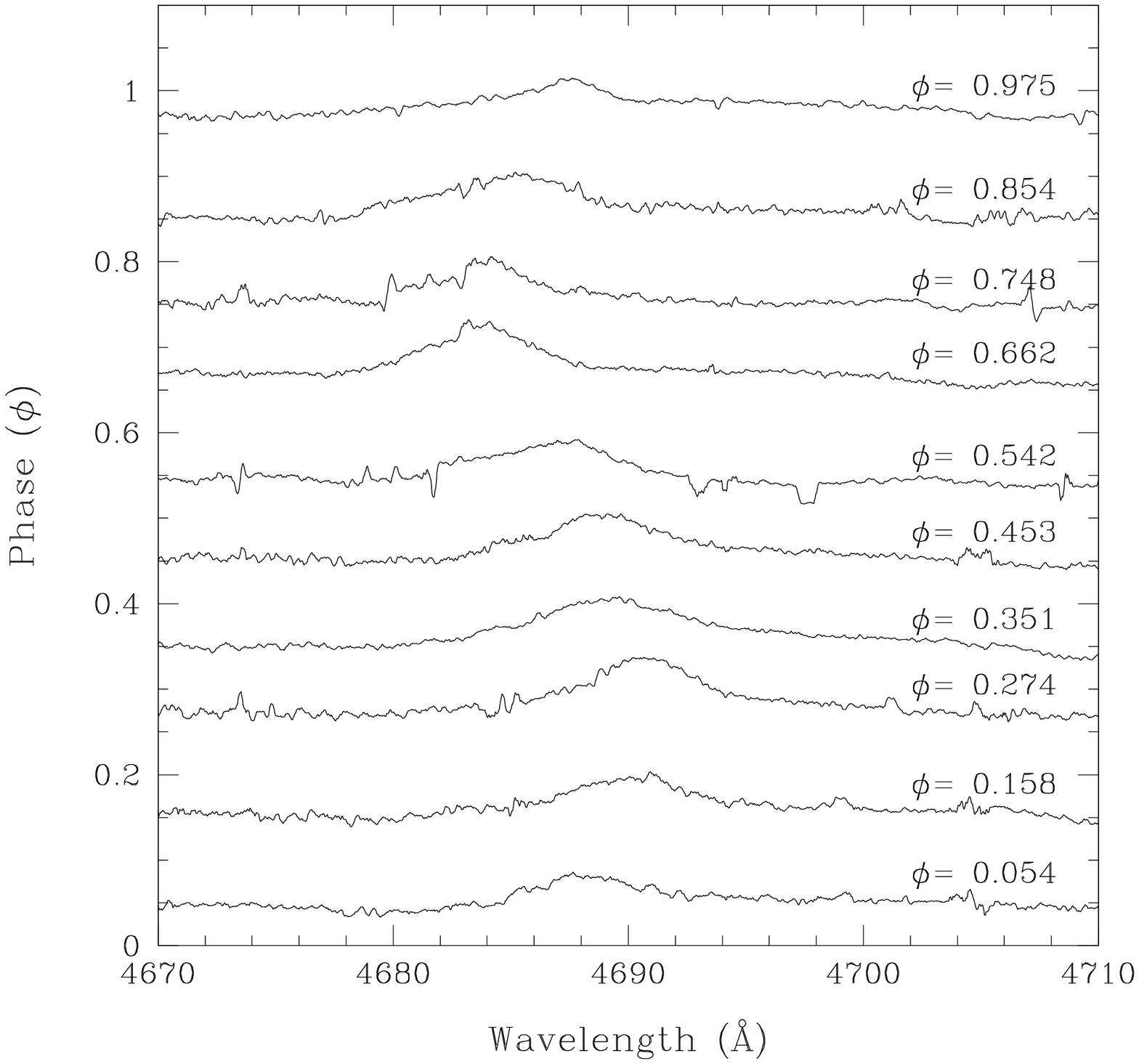}
\caption{{\it Left}: The observed \ion{He}{ii}~$\lambda{4686}$ line profiles in the HD\,150136 spectrum. {\it Right}: The restored \ion{He}{ii}~$\lambda{4686}$ line profiles in the HD\,150136 spectrum when corrected for the absorption lines.}\label{prof_tomo}
\centering
\end{figure*}

The O3V((f$^{*}$)) -- O3.5V((f$^{+}$)) $+$ O5.5--6V((f)) inner system with its small separation between both components likely favours interactions between the stellar winds. If we look closer at the observed spectra, we clearly see that HD\,150136 displays weak emissions in the \ion{He}{ii}~$\lambda{4686}$ and H$\alpha$ lines. To better represent the emissivity of the inner system and notably the possible interactions between the wind of the primary and that of the secondary star, we use a Doppler tomography technique on those line profiles. This technique maps the formation region of these lines in velocity space. The absorption line profiles of the primary and the secondary as well as the presence of the third component in the spectra do not simplify the representation of the emissivity in the inner system (see left panel in Fig.~\ref{prof_tomo}). Therefore, a preliminary treatment to eliminate these contributions in the observed spectra must first be applied. To realize this "cleaning" process, the disentangled spectrum of the third star, shifted by its radial velocity, is subtracted from all the spectra in our dataset. We also take the \ion{He}{ii}~$\lambda{4686}$ line profiles of the synthetic spectra of the primary and the secondary obtained with the CMFGEN code (Sect.~\ref{s_spec}). We thus subtract them by shifting them with their respective radial velocities to remove the absorption parts of the lines. The resulting restored line profiles only take the emission part of the \ion{He}{ii}~$\lambda{4686}$ line into account (Fig.~\ref{prof_tomo}, right panel). However, such a process must be used with caution, especially for the H$\alpha$ line. This line is indeed particularly dependent on the mass-loss rate and on the terminal velocity of the star. Therefore, this cleaning process requires fully reliable absorption line profiles, which is not the case for our three stars. The reconstructed emissions of the \ion{He}{ii}~$\lambda{4686}$ and H$\alpha$ lines are nearly symmetric and follow the orbital motion of the primary component. To analyse their entire profiles, we apply a Doppler tomography technique to these two lines.

The Doppler tomography technique assumes that the radial velocity of any gas flow is stationary in the rotating frame of reference of the system. The velocity of a gas parcel, as seen by the observer, can be expressed by a so-called ``S-wave'' function,
\begin{equation*}
v(\phi) = -v_x \cos (2\pi \phi) + v_y \sin (2\pi \phi) + v_z,
\end{equation*}
where $\phi$ represents the orbital phase, $(v_x, v_y)$ are the velocity coordinates of the gas flow and $v_z$ is the systemic velocity of the emission line. This relation assumes an $x$-axis situated between the stars, from the primary to the secondary, and a $y$-axis pointing in the same direction as the orbital motion of the secondary star \citep[see e.g., ][]{rau02,lin08}. Usually, the method of the Fourier-filtered back-projection based on Radon's transform is used in astrophysics. However, this technique has the disadvantage of generating ``spikes'' in the tomogram if the observations do not densely sample the entire orbital cycle. To limit these artefacts, we thus apply another Doppler tomography program based on the resolution of the matricial system $Af=g$ where $A$ is the projection matrix, $f$ is the vector of emissivity, and $g$ is a vector concatenating the observed spectra. The elements of the projection matrix are equal to 1 if the projection of the S-wave onto the $(v_x, v_y)$ plane is verified or to 0 otherwise. The solutions of the $Af=g$ system are computed from an iterative process called {\it SIRT} \citep[Simultaneous Iterative Reconstruction Technique,][]{kakslaney2001}. This technique starts from an initialisation of the solution vector and computes a correction at each iteration that improves the solution of the previous iteration. These corrections are given by 
\begin{equation*}
f^{(n+1)} = f^{(n)} + \frac{\mu}{||A||^2} A^t \left(g-Af^{(n)}\right)
\end{equation*}
where $\mu$ is a constant so that $0<\mu<2$ and $||\cdot||$ is the matricial norm. This method converges to a least-square solution but requires more computational time than the Fourier-filtered back-projection (for more information on that technique, see \citealt{mahyPhD2011}).

\begin{figure}[htbp]
\centering
\includegraphics[width=8cm,bb=13 10 433 393,clip]{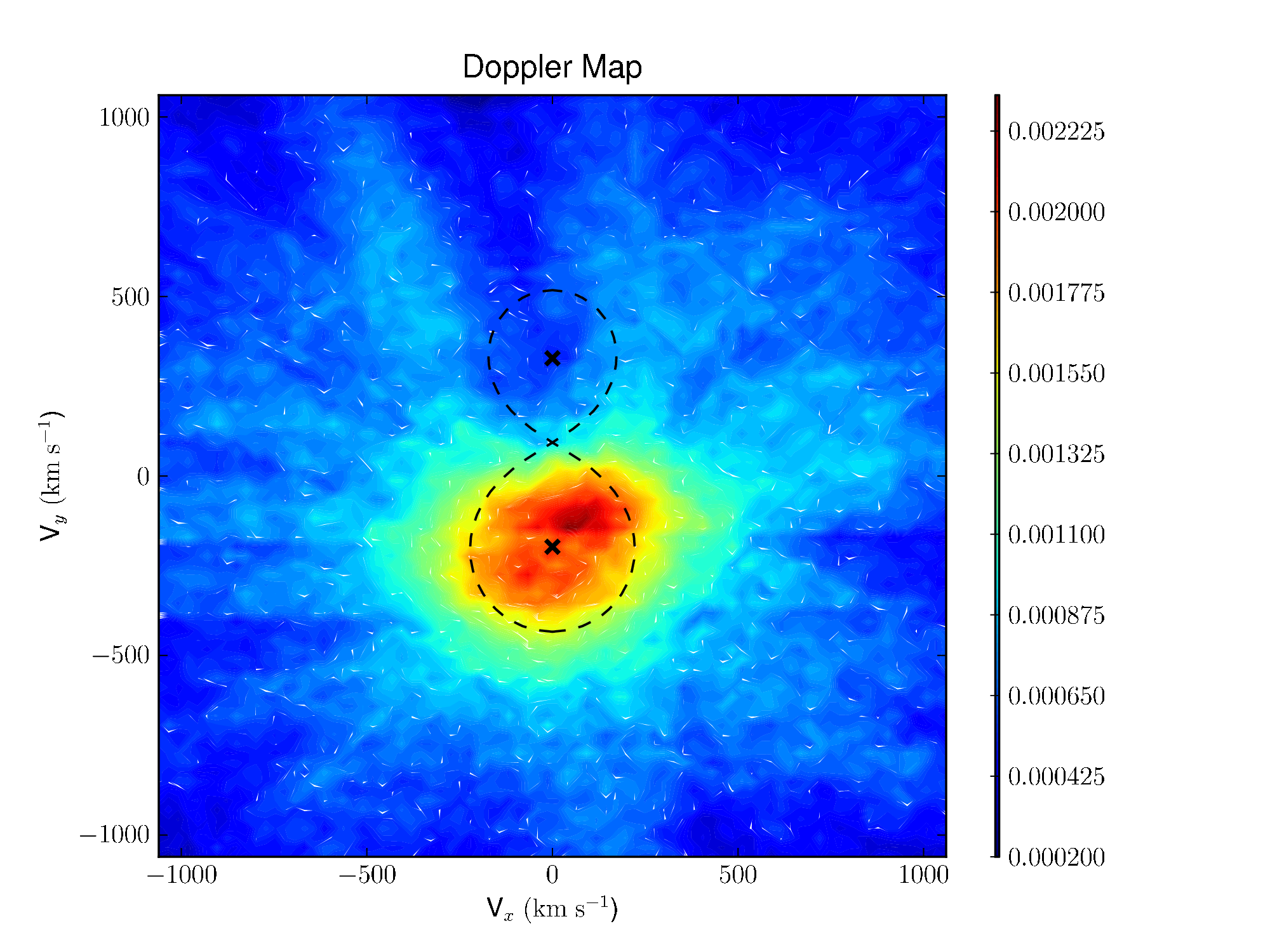}
\includegraphics[width=8cm,bb=13 10 433 393,clip]{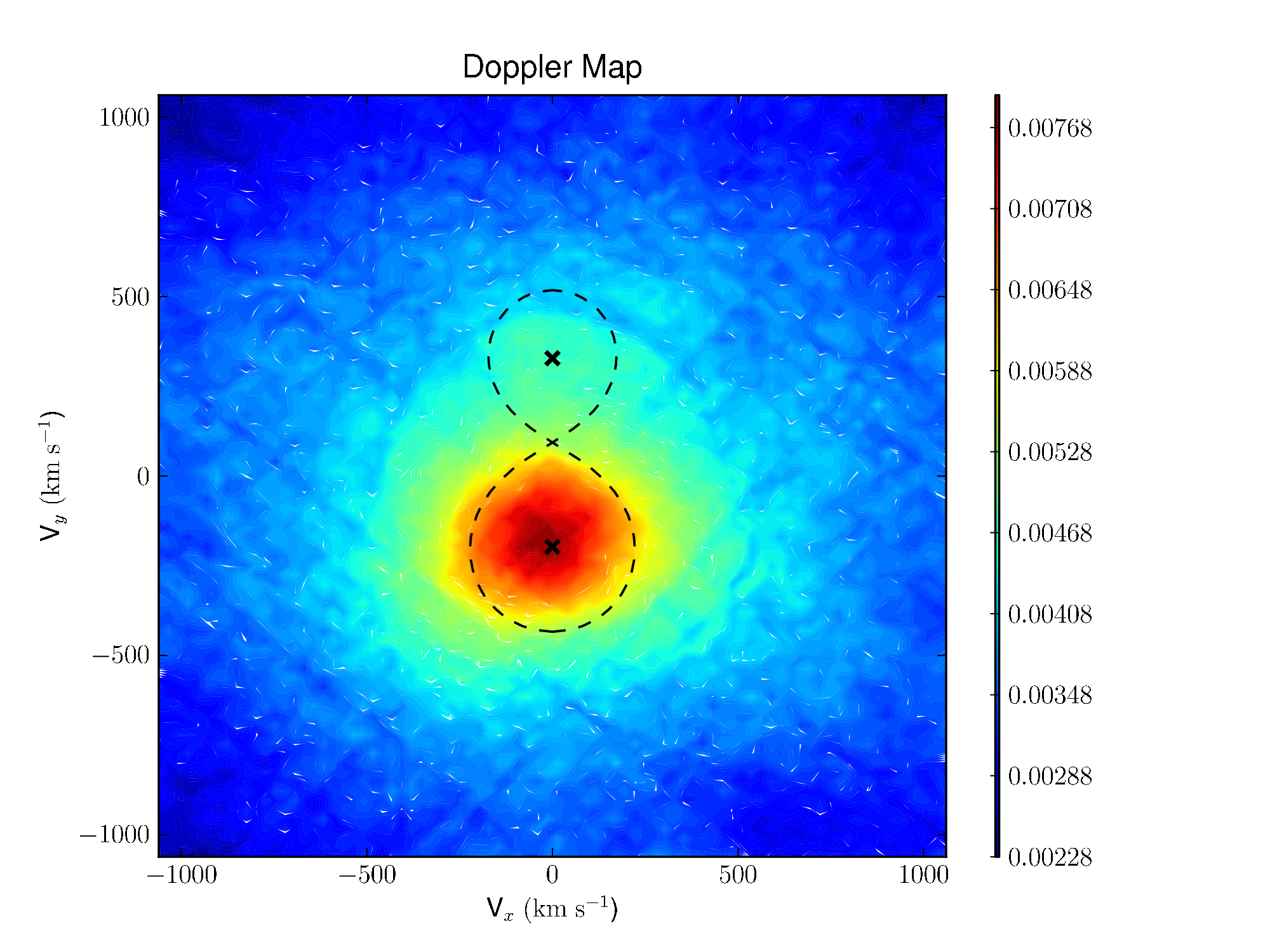}
\caption{{\it Top}: Doppler map of the \ion{He}{ii}~$\lambda{4686}$ line of HD\,150136 computed after removing the absorption line profiles of the primary and secondary stars, as well as of the third component. The crosses correspond to the radial velocity amplitudes of the primary and secondary stars. The shape of the Roche lobe in velocity space (thick dashed line) was calculated for a mass ratio (primary/secondary) of 1.6. The Doppler map was computed with $v_z$ = 0~\kms. On the colourscale, red (blue) indicates the maximum (minimum) emissivity. {\it Bottom}: Same as for the upper panel, but for H$\alpha$.}\label{fig_tomo}
\centering
\end{figure}

The tomograms are computed from multi-epoch spectra. Consequently, we have to correct the radial velocities to bring the spectra to the same systemic velocity on the entire campaign. The \ion{He}{ii}~$\lambda{4686}$ resulting Doppler map (Fig.~\ref{fig_tomo}) exhibits a single region of emission concentrated near the velocity of the centre of mass of the primary star. The maximum of emissivity is located at $(v_x, v_y) \sim (50, -137)$~\kms. For the H$\alpha$ line, we see a similar pattern as for the \ion{He}{ii}~$\lambda{4686}$ line. However, the analysis of the S-wave of the highest emission peak reveals a feature at $(v_x, v_y) \sim (-24, -155)$~\kms. These resulting tomograms do not show any structures reminiscent of a wind-wind interaction or of a Roche lobe overflow. Although the emission peaks are not clearly centred on the centre of mass of the primary, we can assume given the size of the stars that the emissions of the \ion{He}{ii}~$\lambda{4686}$ and H$\alpha$ lines appear to be formed in an unperturbed part of the primary wind. This result coupled with the P\,Cygni profiles obtained from the disentangling thus suggests that the wind of the primary star could be relatively strong for its spectral type. We notice that these Doppler maps are close to those computed by \citet{rau02} for HDE\,228766.


\section{Discussion}
\label{s_disc}

\subsection{The inner system}\label{8.1}

The projected rotational velocities and the radii quoted in Table~\ref{tab_mod} suggest that both components of the inner system of HD\,150136 are in synchronous rotation, i.e., the angular velocity is similar for both stars. Moreover, from the minimum and evolutionary masses reported in Tables~\ref{tab_orb}~and~\ref{tab_mod}, we derive an inclination of about $47.2\degr_{-2.5}^{+2.9}$ for the primary and of about $51.3\degr_{-3.2}^{+1.7}$ for the secondary. Consequently, by assuming that our stellar parameters are close to the reality, we estimate the inclination of the inner system to be about $49\degr \pm 5\degr$. This value favours a non-eclipsing system and agrees with the previous inclination of about $50\degr$ given by \citet{ng05}. We then focus on the possible synchronous co-rotation of the inner system. From the orbital period and the radii, co-rotational velocities of about $276_{-42}^{+57}$~\kms\ and $187_{-30}^{+36}$~\kms\ for the primary and the secondary components, respectively, are found. Such velocities require an inclination of about $38\degr_{-11}^{+16}$ and $47\degr_{-15}^{+37}$ to agree with the measured \vsini. These inclinations are close to the value estimated from the masses. Therefore, we may conclude that the inner system is probably in synchronous co-rotation. 

By adopting a mass ratio of $1.6$ and an $a \sin i = 28.9~\rsun$, we estimate the minimal radii of the Roche lobes of the primary and the secondary from the expression given by \citet{egg83}. We find $(R_{\mathrm{RL}} \sin i )_{\mathrm{P}}= 12.1~\rsun$ and $(R_{\mathrm{RL}} \sin i )_{\mathrm{S}}= 9.8~\rsun$. If this result confirms that the secondary does not fill its Roche lobe, the case of the primary depends on the inclination of the system. By assuming an inclination of $49\degr \pm 5\degr$ and by taking the radius given in Table~\ref{tab_mod}, this component would not fill its Roche lobe either. 
   
\subsection{The tertiary component}\label{8.2}

Previous investigations such as that of \citet{ng05} or \citet{ben06}, mentioned the existence of a third component in HD\,150136. Sometimes, the close visual companion located at about $1.6\arcsec$ \citep{mas98} was quoted as this tertiary star. However, from the present analysis, we conclude that this close visual star cannot be this third component revealed by our analysis. First of all, the spectral classification of the close visual companion is estimated to be later than B3, while we clearly report here an O6.5--7V((f)) classification. Moreover, if we assume an orbital period in the range of $8$ to $15$ years for the outer orbit, the possible $a\,\sin i$ separation between the third component and the inner system is estimated between 3500 and 9100 \rsun\ i.e., between about 16 and 43 AU. With such a separation and assuming a distance of 1.3~kpc for HD\,150136, we calculate a typical angular separation ($a \cos i$) between 14 and 38 mas, much smaller than $1.6\arcsec$ as estimated for the close visual companion reported by \citet{mas98}. On the basis of the spatial separation associated with a mass ratio between the inner system and the tertiary component of about 2.9, this third component should be detectable by interferometry according to the scheme given by \citet{sanaevans2010}. An observation made with SAM/NaCo (Nasmyth Adaptive Optics System, Paranal, Chile) in March 2011 allowed detecting a possible object at the limit of being significant (Sana et al. in preparation). This object would be located below 25~mas. Of course, this information must be taken with great care and needs to be confirmed by future observing campaigns.

Although HD\,150136 is now established as a triple system, it could still have a higher multiplicity. Indeed, the possible close visual component located at about $1.6\arcsec$ could be also bound to the system. In order to determine whether HD\,150136 is a dynamically stable hierarchical triple system, we use the stability criterion quoted by \citet{tokovinin2004}:
\begin{equation*}
P_{\mathrm{out}}(1-e_{\mathrm{out}})^3>5P_{\mathrm{in}}
\end{equation*}
where $P_{\mathrm{in}}$ and $P_{\mathrm{out}}$ are the orbital periods of the inner and the outer systems, respectively, and $e_{\mathrm{out}}$ is the eccentricity of the outer orbit. Even though we do not know precisely the exact values of the parameters of the external orbit, we can still report, in the worst case, that this criterion is satisfied for the configuration of HD\,150136.

\subsection{Evolutionary status}\label{8.3}

\begin{figure}[htbp]
\centering
\includegraphics[width=8cm,bb=38 183 568 694,clip]{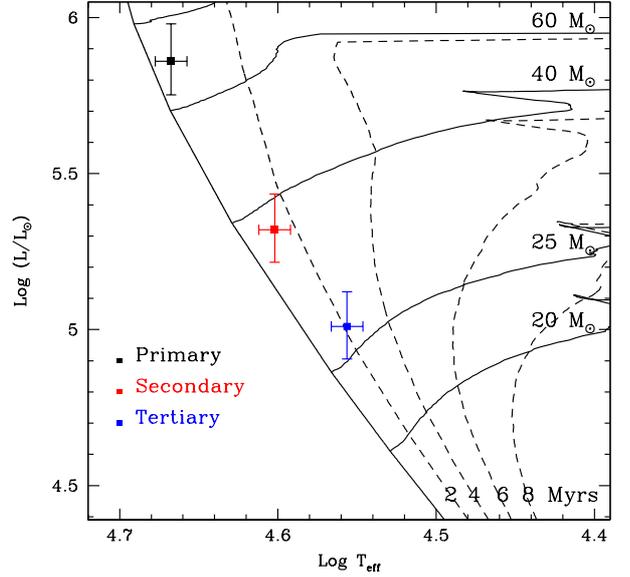}
\caption{HR diagram reporting the position of the three components of HD\,150136. The evolutionary tracks computed for an initial rotational velocity of about 300~\kms, and the isochrones are from \citet{mm03}.}\label{fig_evol}
\centering
\end{figure}

The stellar parameters determined with the CMFGEN atmosphere code allow us to give the positions of the three stars in the Hertzsprung-Russell (HR) diagram. As shown in Fig.~\ref{fig_evol}, all these components are located relatively close to the ZAMS. The isochrones reveal an age of about $0-2$ Myr for the primary and the secondary stars, whilst the tertiary would be slightly older with an age of about $1-3$ Myr. Even though this discrepancy is barely significant, such a difference in age was already observed in several studies. Indeed, \citet[][and references therein]{mar11b} reported that a possible reason to see this difference in the inferred age could be attributed to a different rotational rate between the components. By taking an inclination of $49\degr$ into account for the primary and the secondary component and by assuming that the tertiary is in the same orbital plane as the inner binary, we measure rotational velocities of about 227, 180, and 95~\kms\ for the primary, the secondary, and the tertiary components, respectively. To represent them together, we use the evolutionary tracks computed with an initial rotational velocity of 300~\kms\ (Fig.~\ref{fig_evol}). However, evolutionary tracks based on the individual rotational rates are also investigated, but this does not change the conclusions: the discrepancy does not seem to come from a different rotational rate. It has to be stressed that the evolution of the rotational velocity, hence the efficiency of rotational mixing, is probably quite different in a close binary system where the stars corotate synchronously, on the one hand, and in a rather isolated third star, on the other.

This discrepancy in age is strengthened by determining the stellar parameters (see Sect.~\ref{s_spec}). Even though the parameters of the primary and the secondary stars agree with those of stars on the main sequence, we observe for the third component that its luminosity corresponds to a main-sequence star, whilst its gravity is more reminiscent of an evolved star. However, \logg\ is determined on the basis of the wings of Balmer lines, and its estimate could be wrong if the wings of these lines are poorly reproduced by the disentangling process. In the case of HD\,150136, we see that, due to the broadness of the Balmer lines, the spectral separation between the different components is too small to sample the full width of these lines. Therefore, the deduced value of the \logg\ could be affected by the disentangling process.

Moreover, the parameters of the third star also show an evolutionary mass that is higher than the spectroscopic one, even though both values are lower than what the mass ratio between the inner system and the third star seems to indicate. We notice that the spectroscopic mass is derived from the radius and the \logg\ of the star. Consequently, a wrong estimate of \logg\ affects the determination of the spectroscopic mass. If we consider the evolutionary mass of the inner system (primary and secondary), we find a total mass of about 100~\msun. By adopting the mass ratio we have derived between the inner system and the third star, we obtain a mass of about 35~\msun\ for the tertiary component. This value is close to the evolutionary mass, which would show that the spectroscopic \logg\ would be wrong. This dynamical mass would also lead to a \logg\ value of about 4.1, which is more reminiscent of a main-sequence star. 

Finally, if we assume that the disentangling does not affect the true value of the \logg, we have to look for the real phenomena that could explain such a value. From the different evolutionary schemes of \citet{eggleton1996}, it is however unlikely that the third star has a giant luminosity class. Indeed, according to these authors, a massive triple system such as HD\,150136, i.e., with $M_{\mathrm{P}} > M_{\mathrm{S}} > M_{\mathrm{T}}$, should first undergo a Roche lobe overflow in the inner binary. Since the primary star is more massive than the other ones, it should be the first to evolve, thus implying the mass transfer in the inner system. Therefore, the tertiary component could not be evolved, whilst the primary and the secondary still remain on the main-sequence band. Furthermore, the absence of modified surface chemical abundances and the low \vsini\ derived for the third component also do not support an alternative scenario where the third component would have been captured by the inner system after having been ejected by a supernova kick or by dynamical interactions in the neighbourhood of HD\,150136. All in all, it thus appears that the tertiary component and the two other stars belong to the main sequence and that the most probable cause of a wrong estimate of \logg\ is the disentangling process.  

\subsection{Colliding wind and non-thermal radio emission}\label{8.4}

The presence of an O3V((f$^{*}$))--O3.5V((f$^{+}$)) star in the inner system seems to be mainly responsible for the variations observed in the \ion{He}{ii}~$\lambda{4686}$ and H$\alpha$ lines. However, the proximity of the O5.5--6V((f)) secondary could imply a wind-wind interaction zone between the two stars, even though the disentangling and the Doppler tomography render that unlikely. According to the formalism of \citet{ste92} and the parameters given in Table~\ref{tab_mod}, we expect that the powerful wind of the primary star crashes onto the surface of the secondary star. This scenario is in line with the finding of \citet{ski05}. The collision of the primary wind with the secondary surface could generate an X-ray emission associated with the secondary inner surface. The short timescale X-ray variability could then result from occultation effects very much like those observed for CPD$-41\degr$\,7742 \citep{san05}. Though this scenario seems to agree with the results of the Doppler tomography where we have seen that the emission is only due to the primary star, the position of the contact zone remains unknown because the wind parameters of the secondary star are uncertain. Therefore, we use the mass-loss rate of about $\dot{M} = 10^{-6} \msun$~yr$^{-1}$ derived by \citet{how96} as a function of stellar luminosity, and we assume an inclination of $49\degr$ for the system. These assumptions lead to a probable radiative primary wind, with a cooling parameter $\chi \cong 0.87$. On the basis of different values for the fundamental parameters, \citet{ski05} derived $\chi = 0.2$, also in agreement with a non-adiabatic regime for the X-ray emission from the inner wind-wind interaction region. On the other hand, any X-ray contribution coming from the colliding wind region located between the inner system and the third star should a priori be produced in the adiabatic regime, at least in a very large fraction of the long eccentric orbit.

\begin{figure}
\begin{center}
\includegraphics[width=85mm]{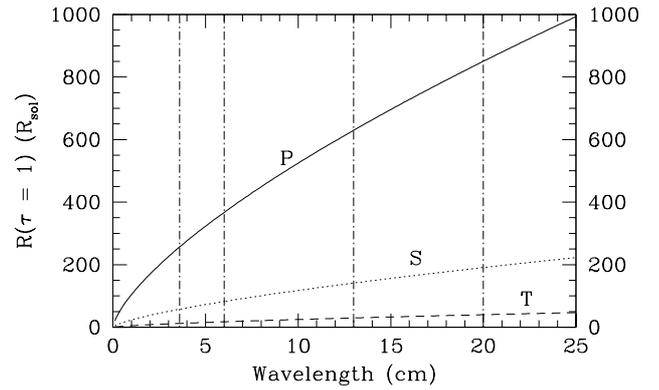} 
\caption{Dependence of the radius of the $\tau = 1$ radio photosphere as
a function of wavelength for the primary (P), the secondary (S) and the
tertiary (T) components. The vertical lines are located at 3.6, 6, 13,
and 20\,cm corresponding to the wavelengths investigated by
\citet{ben06}. We stress that the separation between P and S is estimated at $a\sin{i}=28.9~\rsun$ whilst the separation between (P+S) and T is between 3500 and 9100~$\rsun$.\label{radiophot}}
\end{center}
\end{figure}

The confirmation that HD\,150136 includes a third star is worth
discussing in the context of its non-thermal radio emission. The stellar
wind material is quite opaque to radio photons, which are unlikely to
escape if the absorbing column along the line-of-sight is too thick. The
radius of the $\tau = 1$ radio photosphere can be estimated on the basis
of the formula given by \citet{wrightbarlow75} and \citet{leitherer1995},
following the same approach as \citet{deb04} in the case of the
candidate binary system HD\,168112. Using the stellar wind parameters
given in Table~\ref{tab_mod}, we calculate the radio photosphere radii,
respectively for the primary, the secondary, and the tertiary components.
These radii are plotted as a function of wavelength in Fig.~\ref{radiophot}. 
When discussing the radio emission from massive stars,
one should keep in mind that stellar winds are radio emitters, but in
the thermal regime. On the other hand, the wind-wind interaction in
massive binaries could produce non-thermal radiation in the form of
synchrotron radiation. \citet{ben06} reported on a clearly
non-thermal spectral index between 3.6 and 6 \,cm, but the spectral index
at longer wavelengths (13 and 20\,cm) is more typical of a thermal
spectrum. We therefore focus on the values derived at shorter
wavelengths.
While we are dealing with a triple system, let us separately consider two
possible wind interaction regions: a first one within the short period
sub-system (hereafter, SPC for short period collision) and a second one
involving the third star (hereafter, LPC for long period collision). We
can easily check whether any putative non-thermal radio emission from
the SPC could contribute to the measured non-thermal radio flux. The
curves in Fig.\,\ref{radiophot} are dominated by the primary
contribution, which is easily explained by its high mass-loss rate. The
radii at 3.6 and 6\,cm are of the order of 260 and 370\,\rsun\ for
the primary. The projected stellar separation (a$_1 \sin i$ + a$_2 \sin i$)
is of about 28.9\,\rsun, which translates into a stellar separation of
the order of 38\,\rsun\ for an inclination of about 49$^\circ$ (see
Sect.\,\ref{8.1}). The radio photospheres at 3.6 and 6\,cm overwhelm
thus completely the close primary + secondary system, and consequently
the SPC is deeply buried in the opaque wind. This fact clearly rejects
the possibility that the non-thermal radio photons come from the
interaction between the primary and secondary components. Alternatively,
they could come from the LPC. The radii of the radio photosphere at 6
\,cm constitute therefore minimum boundaries for the distance between
the LPC and, respectively, the primary (solid curve in Fig.~\ref{radiophot})
 and the tertiary (long-dashed curve). The sum of these
two radii (i.e. about 390\,\rsun) should be considered as a lower
limit for the stellar separation between these two stars. That value is
indeed much lower than the stellar separation estimated in Sect.\,
\ref{8.2} on the basis of the probable period of the wide system. A more
stringent lower limit could be derived using values at 20\,cm, since the
free-free absorption increases as a function of wavelength. However, the
non-thermal nature of the radio photons at that wavelength is not
guaranteed by the spectral index reported by \citet{ben06}. We
note that the radio photospheric radii estimated here intimately
depend on the assumed stellar wind parameters, which should be
considered with caution. In particular, the mass loss rate of the
primary may be significantly underestimated here. However, our conclusion relies
on conservative margins.

\section{Conclusion}
\label{s_conc}

The analysis of high-resolution, high signal-to-noise data, collected over several epochs with FEROS, has allowed us to detect a third component physically bound to the inner system for the first time in the spectrum of HD\,150136. In addition, we have revised the spectral classification of the three components to an O3V((f$^{*}$))--3.5V((f$^{+}$)) primary, an O5.5--6V((f)) secondary and an O6.5--7V((f)) tertiary. We also estimated their evolutionary masses as equal to 71, 36, and 27~\msun, respectively. 

We refined the orbital solution of the inner binary system by fitting the radial velocities of the different components of the SB3 system together. The results have improved the accuracy of the majority of the parameters reported by \citet{ng05}. Moreover, we found that the inner system is probably in synchronous co-rotation. The agreement between the evolutionary and the Keplerian masses would suggest a probable inclination of about $49\degr$, implying dynamical masses of about 64, 40, and 35~\msun. 

From the systemic velocities measured on each epoch, we estimated that the orbit of the third star should have a period between 2950 and 5500~d and an eccentricity higher than 0.3. More data are necessary, however, to achieve an accurate orbital solution of the outer orbit. A possible alternative would be to observe HD\,150136 by interferometry to constrain the missing orbital parameters.

The determination of the individual stellar parameters of the three stars has yielded a luminosity and a gravity corresponding to main-sequence stars for the primary and the secondary components, but these parameters are trickier for the third star. Indeed, we observed a clear discrepancy between its luminosity (comparable to main-sequence stars) and its $\log g$ (typical of giant stars). The origin of such a difference is thus unclear, but it seems that it could come from the disentangling procedure, notably because the spectral separation is not sufficient to sample the full width of the broadest lines, such as the Balmer lines, for the three components. Consequently, HD\,150136 is probably composed of at least three O-type stars located on the main sequence and with inferred ages between $0-3$ Myr. Finally, we showed that the non-thermal radio emission should come from the wind-wind interaction region between the third star and the inner short-period system, region that should be the site for particule acceleration in this multiple system.

\begin{acknowledgements}
This research was supported by the PRODEX XMM/Integral contract (Belspo), the Fonds National de la Recherche Scientifique (F.N.R.S.) and the Communaut\'e fran\c caise de Belgique -- Action de recherche concert\'ee -- A.R.C. -- Acad\'emie Wallonie-Europe. The authors thank Thomas Fauchez for a first detailed look to the data in the framework of a student work. We also thank Dr. F. Martins for his helpful comments and Pr. D.J. Hillier for making his code CMFGEN available. We are grateful to the staff of La Silla ESO Observatory for their technical support.
\end{acknowledgements}

\bibliography{laurent.bib}

\end{document}